%% file: main.tex
\pdfoutput=1
\documentclass[11pt]{article}
\usepackage{authblk}
\usepackage[]{acl2023}
\usepackage{times}
\usepackage{xcolor}
\usepackage{amsmath}
\usepackage[thinlines]{easytable}
\usepackage{graphicx}
\usepackage{latexsym}

\usepackage{array}
\usepackage{booktabs}
\usepackage{pgfplots}
\usepackage{algorithm}
\usepackage{lipsum}
\usepackage{mathtools}
\usepackage{cuted}
\usepackage{makecell}
\usepackage{algpseudocode}
\usepackage{graphicx}
\usepackage[T1]{fontenc}
\usepackage[utf8]{inputenc}
\usepackage{microtype}
\usepackage{xcolor}
\usepackage{soul}
\usepackage{fullpage}
\usepackage{enumitem}
\usepackage{pgfplots}
\usepackage{algorithm}
\usepackage{algpseudocode}
\usepackage{graphicx}
\usepackage{placeins}
\usepackage{tabularx}
\usepackage{makecell}
\usepackage{booktabs}       
\usepackage{array}         
\usepackage{colortbl}
\usepackage{times}
\usepackage{graphicx}
\usepackage{latexsym}
\usepackage{array}
\usepackage{booktabs}
\usepackage{pgfplots}
\usepackage{algorithm}
\usepackage{algpseudocode}
\usepackage{graphicx}
\usepackage[T1]{fontenc}
\usepackage[utf8]{inputenc}
\usepackage{microtype}
\usepackage{xcolor}
\usepackage{soul}
\newcommand{\Lagr}{\mathcal{L}}
\title{Noise-Robust Dense Retrieval via Contrastive Alignment Post Training}
\author[1,]{Daniel Campos}
\author[1]{ChengXiang Zhai}
\author[2]{Alessandro Magnani}
\affil[1]{Department of Computer Science, the University of Illinois Urbana-Champaign}
\affil[2]{Walmart Labs}
\begin{document}
\maketitle

\begin{abstract}
The success of contextual word representations and advances in neural information retrieval have made dense vector-based retrieval a standard approach for passage and document ranking. While effective and efficient, dual-encoders are brittle to variations in query distributions and noisy queries. Data augmentation can make models more robust but introduces overhead to training set generation and requires retraining and index regeneration. We present Contrastive Alignment POst Training (CAPOT), a highly efficient finetuning method that improves model robustness without requiring index regeneration, the training set optimization, or alteration. CAPOT enables robust retrieval by freezing the document encoder while the query encoder learns to align noisy queries with their unaltered root. We evaluate CAPOT noisy variants of MSMARCO, Natural Questions, and Trivia QA passage retrieval, finding CAPOT has a similar impact as data augmentation with none of its overhead.
\end{abstract}
\section{Introduction}
\label{sec:intro}
\input{intro}
\section{Related Work}
\label{sec:rel}
\input{related}
\section{Query Encoders Meet Noise}
\label{sec:create}
\input{method}
\section{Incorporating Noise By Aligning Representations}
\input{experiment}
\section{Discussion}
\input{discussion}
\section{Conclusion and Future Work}
\input{conclusion}
\bibliographystyle{ACM-Reference-Format}
\bibliography{references}
\appendix
\input{appendix}
\end{document}

%% file: intro.tex
Contextual language representations derived from Large Language Models (LLM) have led to impressive improvements in performance across many tasks in natural language processing, such as sentiment analysis, topic detection, and question answering. \\
In information retrieval, LLM based cross encoders and bi-encoder models have led to major improvements in relevance on benchmarking datasets like MSMARCO \cite{nguyen2016ms} and Natural Questions \cite{Kwiatkowski2019NaturalQA} and have been adopted as common backbones for many industrial deployments in search. Unlike traditional term-based search, contextual representations excel at semantic search, which can improve relevance as it matches intent instead of keywords. \\
\begin{figure}
    {\scalebox{0.45}{\includegraphics[width=\textwidth]{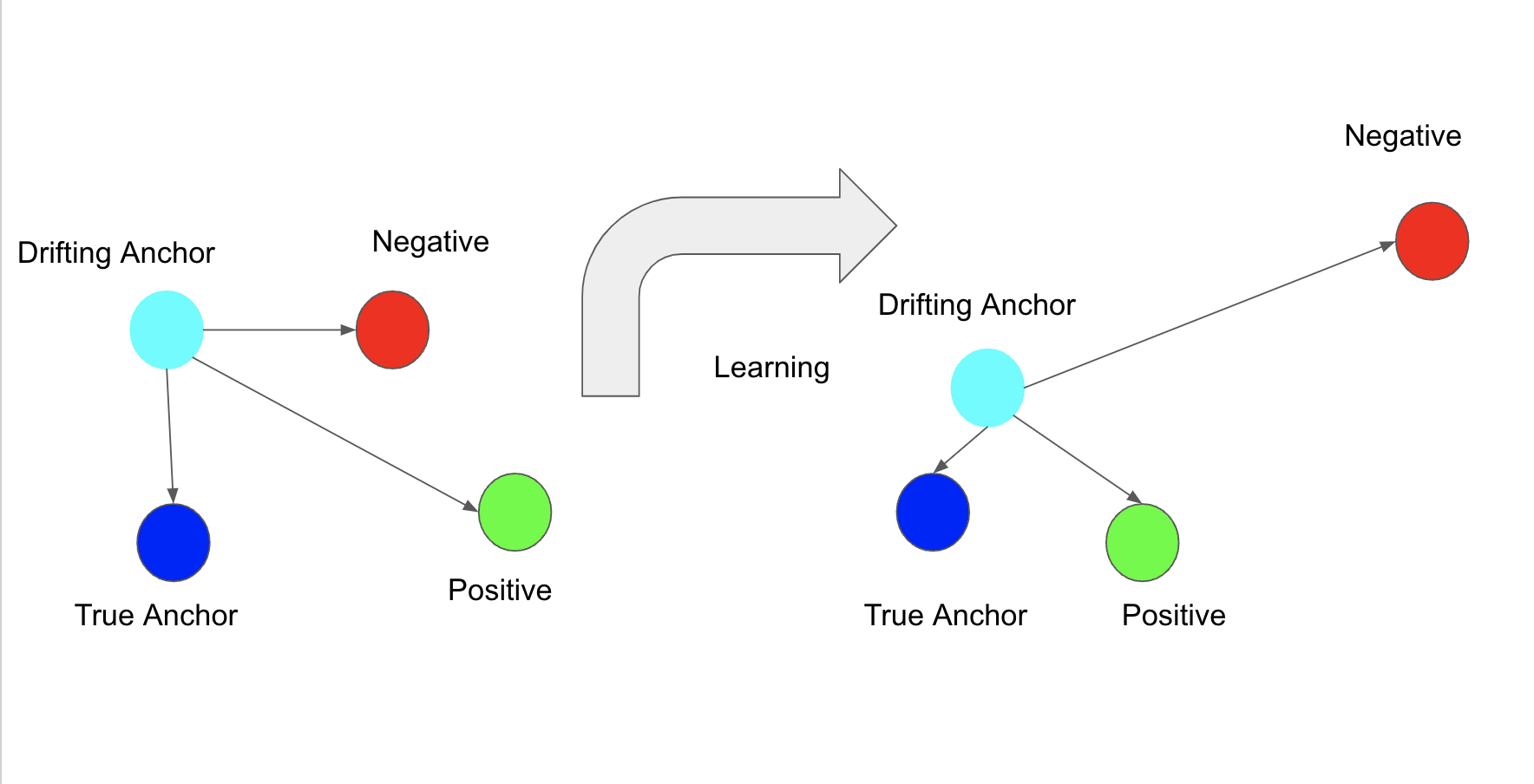}}}
  \caption{To learn a align the representation of queries and their counterparts with noise, a contrastive loss is used. Since the document encoder and its relative relation to the query encoder are frozen, an anchoring vector keeps the aligned encoder from drifting from its original learned representation.}
  \label{fig:fig1}
\end{figure}
While neural methods excel on well-formulated academic benchmarks, performance falters when faced with domain shifts or queries with misspellings or typos. On recent benchmarks like BEIR \cite{Thakur2021BEIRAH}, cross-encoders are more robust to shifts in domain than bi-encoders. \\
When looking at queries with some noise like typos and misspellings, the same holds \cite{zhuang-zuccon-2021-dealing}. Despite their superior performance on noisy queries and domain shifts, cross-encoder inference requirements make them too expensive to use at scale. \\
To avoid inference inefficiency of cross-encoder, bi-encoders have emerged as a popular method for retrieval, particularly for candidate generation in multi-stage ranking systems. Bi-encoders leverage query and document models trained to match their representations in a latent space. Since document representations are query independent, they only need to be generated once, and the inference load is limited to a single run of the query encoder. \\
Research on bi-encoder has been driven by the availability of large training datasets such as MSMARCO \cite{Campos2016MSMA}, Natural Questions (NQ)\cite{Kwiatkowski2019NaturalQA}, and Trivia QA (TQA) \cite{Joshi2017TriviaQAAL}. These datasets have allowed deep explorations on how to improve training procedure \cite{Qu2021RocketQAAO}, decrease index size \cite{Yamada2021EfficientPR}, and model efficiency \cite{Khattab2020ColBERTEA}.  Despite the tremendous success, these neural methods models are brittle to subtle search domain shifts and minor query formulation variations\cite{Wu2021AreNR}.  \\
While there has been plenty of work that has shown how neural methods are not robust to typos \cite{Wu2021AreNR} \cite{Penha2021EvaluatingTR} \cite{Sidiropoulos2022OnTI} \cite{Sidiropoulos2022AnalysingTR} \cite{zhuang-zuccon-2021-dealing}  all approaches which improve performance either require a new optimized general model such as CharBERT \cite{Zhuang2022CharacterBERTAS} or require retraining with data augmentation \cite{zhuang-zuccon-2021-dealing}. While effective, both approaches introduce a sizable overhead in dataset generation and augmentation or language model pre-training. Moreover, despite the effectiveness of these two techniques, further study is required to understand the interplay between data augmentation and  curriculum learning \cite{Mao2022CurriculumCC} and topic-aware sampling \cite{Hofsttter2021EfficientlyTA}. \\
\begin{table*}[htb!]
    \scalebox{0.60}{
    \begin{tabular}{|l|l|l|l|}
    \hline
        Noising Function & Alteration Type & Original & Alteration \\ \hline
        Determiner  & Syntactic & who sang waiting for a girl like you & who sang waiting a a girl like you \\ \hline
        Synonym & Semantic & \makecell{Which was the first European country to \\ abolish capital punishment?} & \makecell{Which was the first European country \\ to abolish majuscule punishment?}\\ \hline
        Lemmatize & Syntactic & who plays young dr mallard on ncis & who play young dr mallard on ncis \\ \hline
        Stemming & Syntactic & who recorded the song still the one? & who record the song still the one? \\ \hline
        Random Character Swap (RCS) & Surface & big little lies season 2 how many episodes & big litt e lies season 2 how many episodes \\ \hline
        Keyboard Character Swap (KCS) & Surface & when did veterans day start being called veterans day & when djid veterans day start being called veterans day \\ \hline
        Character Delete (CD) & Surface & when did big air snowboarding become an olympic sport & when did big air snowboarding become an olympic sort \\ \hline
        Reorder Word (RW) & Surface & who is the main character in green eggs and ham & who is the main character and green eggs in ham \\ \hline
        Back-Translation (BT) & Semantic & what is project charter in project management & What is a project charter in project management \\ \hline
        Paraphrasing & Semantic & turkey and china time difference & \makecell{Time difference between Turkey and China in the middle \\ of the night, depending on the time difference.} \\ \hline
    \end{tabular}}
    \caption{Example of the forms of query noise that we leverage to evaluate how robust bi-encoders are to noise.}
    \label{tab:query-noise}
\end{table*}
Seeking to improve the performance of the query encoder on noisy queries with high efficiency possible, we introduce \textbf{C}onstrastive \textbf{A}llignment \textbf{PO}st \textbf{T}raining (CAPOT). To avoid complicated dual encoder training regimes, CAPOT assumes that the document encoder and index are immutable and learn an improved query representation without altering existing relations to the index. As shown in figure \ref{fig:fig1}, CAPOT uses a traditional contrastive loss \cite{Schroff2015FaceNetAU} where queries with noise (positive samples) should be closer to the anchor (query without noise) than unrelated queries. Unlike a traditional contrastive loss, CAPOT introduces a notion of an anchoring loss between the unaltered model and the aligned model. As the model learns to group noisy queries with their unaltered roots, we also constrain its ability to alter the representation aligned with the unaltered document encoder.\\
The main contributions of our work are as follows:
\begin{itemize}
\item We introduce CAPOT, a highly efficient fine-tuning method for improving performance on noisy queries without retraining a model or index regeneration. 
\item We demonstrate that CAPOT is incredibly effective at making the encoder robust, particularly with typos. Using CAPOT approximates the impact of data augmentation without the associated computational overhead.
\item We demonstrate that CAPOT is robust enough to prove functional with completely unsupervised data. Using the ORCAS dataset, CAPOT can improve performance without access to the query training distribution.
\end{itemize}

%% file: related.tex
\textbf{Bi-Encoders}, commonly called dual-encoders or dense retrievers, decompose ranking by leveraging the inner product of query and document representations to produce a relevance score for query document pairs. Since their document representations are query invariant, they can be pre-computed and loaded into an Approximate Nearest Neighbor (ANN) such as FAISS \cite{johnson2019billion}. The $k$ closest documents can be found for each query with minimal latency at run time. Since bi-encoders leverage LLM such as BERT \cite{Devlin2019BERTPO}, they are often limited to ranking short passages of text and are commonly referred to as Dense Passage Retrievers (DPR) \cite{Karpukhin2020DensePR}. Driven by their efficiency in deployment and relevance performance, DPR-based models have rapidly become the building blocks for systems doing product search \cite{Magnani2022SemanticRA}, open domain question answering \cite{Karpukhin2020DensePR} and customer support \cite{Mesquita2022DenseTR}.\\
Recent work has heavily focused on improving the relevance of DPR models by improving the negative sampling using methods like ANCE \cite{Xiong2021ApproximateNN} and in-batch negatives \cite{Lin2021InBatchNF}. While effective DPR models are brittle to shifts in the domain, minor variations can cause a complete collapse in relevance. Li et al. '2022 introduced methods for improving such performance by having a single query encoder leverage multiple document encoders to transfer between domains \cite{Li2022AnEA}. While effective, such a method carries a high computational load as multiple indexes must be maintained and updated. \\
\textbf{Data Augmentation} (DA) is a popular approach for improving how well models perform on new or noisy data. In data augmentation, training is extended by augmenting the training data with modifications or perturbations which match the desired model behavior. DA is extremely common in computer vision where training data is commonly rotated, blurred, cropped, or zoomed-in/out \cite{Mikoajczyk2018DataAF} \cite{Zhong2020RandomED}. \\
DA has become increasingly more popular in NLP and has been used to improve model performance \cite{Jiao2020TinyBERTDB}, simulate large-scale training data when it is not available \cite{Li2020ADD}, and mitigate bias \cite{Lu2020GenderBI} in existing datasets. A detailed survey on DA approaches for NLP has been complied by Feng et al. 21' \cite{Feng2021ASO}.\\
\textbf{Contrastive Learning} builds on the notion of a contrastive loss \cite{Chopra2005LearningAS}, which seeks to create clusters in the embedding space such that examples with a shared class are far from other classes but close to each other. Much like learning that queries with noise have a shared intent, Schroff et al. 15' leverage contrastive learning to recognize faces despite different angles and perspectives \cite{Schroff2015FaceNetAU} by using a triplet loss. This approach is a natural fit for the world of search as relevance is at its core clustering relevant items close together and far from irrelevant items. Recently, contrastive learning has become a method for learning relevance at the corpora scale \cite{Xiong2021ApproximateNN} and improving DPR on noisy queries, \cite{Sidiropoulos2022AnalysingTR} \cite{Chen2022TowardsRD}

%% file: method.tex
\subsection{Generating Noisy Queries}
\label{sec:making-noise}
While previous work has studied the impact of minor variations to queries, such as typos and misspellings, query noise is much more diverse. Seeking to expand this understanding, we explore the impact of query alterations that evaluate surface, syntactic, and semantic alterations. To apply noise to a query, we either edit a query to introduce a specific type of noise or rewrite the query to simulate similarly worded intents. Each query that is altered has a notion of its anchor, either a character, word or a group of words, which is selected where noise is applied. To achieve this for each query, a character or word index is randomly selected. Then, noise is applied to the left, right, or at the noising index (replacing the existing index) with equal probability. Example alterations are in table \ref{tab:query-noise}. \\
To study the impact of surface-level alterations, we introduce noise in queries by simulating misspellings and typos by swapping, eliminating, or shuffling characters in a query. To understand how models respond to typos or character omissions, we delete a character (DC), inject a random character (RCS), or simulate a keyboard-based typo by injecting a character close to its neighbor on a keyboard (KCS). We swap the indexed word with another word in the query to understand how systems may work when faced with natural shifts in keyword queries. \\
To study syntactic alterations, we introduce noise that alters the syntax of the query introducing lemmas, stems, synonyms, and determiners using tools from the NLTK toolkit \cite{bird2009natural}. Synonyms are introduced using NLTK's interface with WordNet \cite{Fellbaum2000WordNetA}, and exact synonyms for a single word are introduced.  Determiners, affixes that occur with nouns and commonly are not discriminative for search, are introduced similarly to typos to the left or right of noun phrases. Lemma's return words to their canonical root while stemming reduced word inflection using the Porter-stemmer. We select up to five words per query to attempt stemming/lemmatization, but many queries do not have any words which can be stemmed or lemmatized versions and, as a result, are un-noised. \\
Exploring semantically similar queries, we leverage paraphrasing, back-translation, and synonyms. To paraphrase, we rewrite queries using a T5 \cite{Raffel2020ExploringTL} sequence-to-sequence model, which has been fine-tuned on the PAWS \cite{Zhang2019PAWSPA} dataset. For back-translation, we use OpenNMT's \cite{klein-etal-2017-opennmt} to translate queries from English to another language and then back to English after exploring performance using German, French, Italian, and Spanish to find the German to have the best quality and use only those. It is worth noting that these semantic noising methods are the most likely to alter the true query intent, as seen by the 'hallucinations' in table \ref{tab:query-noise} paraphrase alteration. \\
Using the aforementioned noising approaches, we noise the queries on the MSMARCO \cite{Campos2016MSMA} \footnote{https://huggingface.co/datasets/spacemanidol/msmarco-passage-query-variation} \footnote{https://huggingface.co/datasets/spacemanidol/rewrite-noisy-queries}, Natural Questions \footnote{https://huggingface.co/datasets/spacemanidol/wikipedia-nq-query-variation} \footnote{https://huggingface.co/datasets/spacemanidol/nq-noising} \cite{Kwiatkowski2019NaturalQA}, and the Trivia QA \cite{Joshi2017TriviaQAAL} \footnote{https://huggingface.co/datasets/spacemanidol/wikipedia-trivia-query-variation} Passage Ranking datasets. \\
\subsection{Baseline Performance}
\begin{figure}[!htb]
\begin{tikzpicture}
\scalebox{0.85}{
\begin{axis}[
    title={Bi-encoder Recall Accuracy by Recall Set size},
    xlabel={Recall Set Size},
    ylabel={Recall Accuracy},
    xmin=20, xmax=200,
    ymin=50 , ymax=95,
    xtick={20, 100, 200},
    ytick={50, 60, 70, 80, 90, 95},
    legend pos=south west,
    ymajorgrids=true,
    grid style=dashed,
    legend style={nodes={scale=0.4, transform shape}}, 
    legend image post style={mark=*}
]
\addplot[
    color=blue,
    mark=square,
    ]
    coordinates {
    (20, 79.73) (100, 85.98) (200, 88.25)
    };
\addplot[
    color=red,
    mark=square,
    ]
    coordinates {
    (20, 71.56) (100,81.58) (200, 83.91)
    };
\addplot[
    color=green,
    mark=triangle,
    ]
    coordinates {
    (20, 71.63) (100, 88.79) (200, 93.78)
    };
    
\addplot[
    color=purple,
    mark=triangle,
    ]
    coordinates {
    (20, 56.65) (100, 74.84) (200, 80.67)
    };
\addplot[
    color=orange,
    mark=star,
    ]
    coordinates {
    (20, 79.40) (100,85.01) (200, 86.66)
    };
\addplot[
    color=black,
    mark=star,
    ]
    coordinates {
    (20, 75.86) (100, 84.72) (200, 84.72)
    };
\legend{NQ, NQ w/noise, MSMARCO, MSMARCO w/noise, TriviaQA, TriviaQA w/noise}
 \end{axis}}
\end{tikzpicture}
    \centering
    \caption{Bi-encoder recall accuracy on noisy and non-noisy queries with variations of recall set size and datasets. }
    \label{fig:baseline-impact-query-noise}
\end{figure}
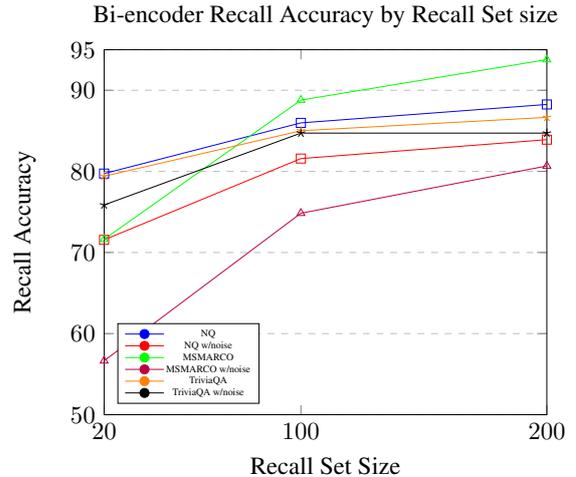
In production workloads, bi-encoders are most commonly used for early retrieval, where the sets they produce are then reranked using a cross-encoder. Given cross-encoder are more robust to typos \cite{Sidiropoulos2022AnalysingTR}, our work focuses exclusively on evaluating the impact of noise on the retrieval accuracy of bi-encoders. \\
To do so, we train  a series of task-specific bi-encoders leveraging the open-source bi-encoder-focused library Tevatron \cite{Gao2022TevatronAE} with task-specific training parameters found in \ref{tab:capot-bi-encoder-hy} on the widely used and studied MSMARCO \cite{Campos2016MSMA}, Natural Questions (NQ) \cite{Kwiatkowski2019NaturalQA} and TriviaQA \cite{Joshi2017TriviaQAAL} passage retrieval datasets. \\
For contextual representations, each encoder uses  a pre-trained BERT \cite{Devlin2019BERTPO} model for its initialization, and we use separate models for the query and document models. Representations are taken from the unaltered 768-dimension vectors based on the last hidden representation of the CLS token. \\
For each dataset, we train each model using 5 different random seeds with fixed optimal hyperparameters, generate seed and task-specific indexes and evaluate the retrieval impact of queries with noise and unaltered routes. We evaluate the impact on retrieval by measuring the impact on retrieval accuracy at $k$ with a $k={20,100,200}$.  \\
As shown in figure \ref{fig:baseline-impact-query-noise} on the impact of averaged noise, our experimental results align with prior research. There is a wide variation of impact as the long, trivia-inspired queries of Trivia QA see minor losses in accuracy compared to the real-world web search queries of MSMARCO, up to 12\% the impact. Besides the impact of query type, we also notice the large role recall set size plays on the relative degradation in retrieval accuracy. Across tasks and datasets, increasing the recall set from 20 to 200 decreases the impact on accuracy by about 50\%. \\ 
Focusing on the impacts of individual types of noise shown in \ref{sec:full-noise-impact}, we see that queries with surface alteration, such as typos, see the largest loss. Despite featuring real-world search engine queries with noise, on MSMARCO, there is nearly a 30\% loss in retrieval accuracy for queries with typos, dropping from 71\% to 41\%. On all datasets, queries with character-level alterations see a 50\% average higher loss in accuracy than other alterations. This large gap can be attributed to the vocabulary construction method of BERT and BERT-like models, where a minor alteration to a single character can produce large variations in tokenization.\\
In the absence of model optimization, data augmentation, or post-training optimization, the clearest way to make dense retrieval robust to noise is to expand the recall set and allow cross-encoders to re-rank the expanded results.\\

%% file: experiment.tex
\begin{tiny}
\begin{equation} 
\label{eq:1} 
    \Lagr_{c}(x, x^+, x^-) = \sum_{x \in X} \max(0,\mathrel{\Vert} f(x) -  f(x^+) \mathrel{\Vert}_2^2 - \mathrel{\Vert} f(x) - f(x^-) \mathrel{\Vert}_2^2 + \epsilon)
\end{equation}\\
\begin{equation} 
\label{eq:2} 
    \Lagr_{a}(x) = \sum_{x \in X} \max(0,\mathrel{\Vert} f(x) - f_a(x) \mathrel{\Vert}_2^2 + \epsilon_{a})
\end{equation}
\begin{equation} 
\label{eq:3} 
    \Lagr_{r}(x^+, x^-) = \sum_{x \in X} \max(0,-y * (f(x^+)-f(-x) + \epsilon_{r})
\end{equation}
\begin{equation} \label{eq:4} 
    \Lagr_{CAPOT}(x, x^+, x^- ) = \sum_{x \in X} \tau_{c} * \Lagr_{c} 
    + \tau_{a} *  \Lagr_{a} + \tau_{r} * \Lagr_{r} 
\end{equation}
\end{tiny}
\subsection{Motivation}
A robust query encoder seeks to represent queries with a shared intent in a common latent space such that minor variations in the formulation of the intent lead to similar document ranking. Prior work has shown that data augmentation and typo-optimized models increase model robustness, but it is not without cost.\\
Data augmentation requires changes to existing training methodologies and complete regeneration of the passage index. Given that the generation of the passage index can take longer than it does to train the model \cite{Karpukhin2020DensePR} regenerating a new index and retraining a model every time a novel form of noise is discovered is not tractable. Optimized pre-trained models can provide effective modeling solutions. However, given the rapid iteration pace of pretrained language models, making typo-aware variants for each new advance is hard to scale.\\
Motivated to improve performance without altering the underlying pretrained model or the bi-encoder training regime, we introduce CAPOT, a new methodology for increasing model robustness which is \textbf{computationally inexpensive} and \textbf{independent of training}. CAPOT works well because it can focus on improving the query encoder and leverages the short nature of queries to scale to large batch sizes. 
\subsection{CAPOT}
\textbf{C}ontrastive \textbf{A}lligment \textbf{Po}st \textbf{T}raining (CAPOT) is an expansion on traditional contrastive learning focused on making dual encoders robust to noise. The goal of CAPOT is to allow representations of noisy queries to be close to their original on the traditional triplet contrastive loss \cite{Schroff2015FaceNetAU} in \ref{eq:1}, where $f$ is a query encoder, $x$ is the original query, $x^+$ is a query where noise has been introduced, and $x^-$ is a negative query selected at random \footnote{We explore the usage of hard negatives mined from nearby query representation but did not find any measurable impact}. We modify \ref{eq:1} to scale the role of positive and negative samples using term specific in$\tau_{positive}$ and $\tau_{negative}$\ parameters. \\
While \ref{eq:1} allows the query-encoder to represent queries and noisy queries in a similar latent space, it has the unwanted side effect of query representation drifting related to the learned notion of relevance. Without controlling for this drift, a complete collapse in ranking accuracy came at the expense of effective representation of noisy samples. To avoid this, we introduce an anchoring term,\ref{eq:2}, that minimizes the drift between learning a notion of relevance and shared embeddings for queries with noise where $f$ is the noise-robust query-encoder, and $f_a$ is a copy of the unaltered frozen query-encoder, optimized for an existing document encoder and document index. \\
Seeking to improve performance further, we add a ranking loss as shown in \ref{eq:3} between the anchored model $f_a$ and $f$ where the model learns that $f(x^+)$ always ranks higher than $f_a(x)$. While this loss component is not crucial, we can leverage this to improve model performance slightly. \ref{eq:1},\ref{eq:2} and \ref{eq:3} are combined to form the CAPOT, \ref{eq:4}. 
\subsection{Experimental Approach}
\label{sec:ea}
To qualify the effectiveness, we explore how alignment can improve performance on noisy queries before and after bi-encoder training and compare them to data augmentation. We then compare the performance of the aligned models with unaltered baselines and models trained with DA. Except for models aligned with CAPOT, each experiment requires a complete training run and index generation, which can be quite slow. Each experiment is performed across 5 seeds, and we use the same evaluation metrics previously discussed and report the mean performance over five seeds.   \\
To quantify the ability of post-training alignment, we take the converged baseline models and apply CAPOT 
to align the model on the training portion of the query set. Once aligned, a model is retrieved on the unaligned, fixed document index generated during our baseline experimentation. Since queries are short and batch sizes can be scaled easily, it's important to note how fast this is. A single 2080ti NVIDIA GPU using CAPOT takes under 60 minutes on the NQ dataset. \\ 
To explore if alignment can happen before training, we leverage the ORCAS \cite{craswell2020orcas} dataset to generate a corpus of 10 million queries. Using these queries, we create positive and negative noisy samples using the same noising approach discussed in \ref{sec:making-noise} making a dataset of 100 million queries called Noisy-ORCAS \footnote{https://huggingface.co/datasets/spacemanidol/CAPOT-queries}. Using these 100m queries, we align the representation of queries and their noisy counterparts using a BERT-base model optimized for masked-language modeling. Given the scale of this dataset, We train for a single epoch on the Noisy-ORCAS corpus using the $\tau_{positive}=1.0$,$\tau_{negative}=0.1$,and $\tau_{anchor}=1.0$ on 4 V100 GPUs with a batch size of 2048. Then, we leverage this optimized model to initialize our unaltered bi-encoder model's training procedure. Then, this model is trained on our datasets and evaluated similarly to the baseline.  We refer to models trained this way as(PT), and each usage of PT requires retraining and index regeneration. 
\subsection{Experimental Results}
\begin{figure}[!htb]
\begin{tikzpicture}
\scalebox{0.85}{
\begin{axis}[
    title={Relative Degradation in Retrieval  vs. Recall Set size},
    xlabel={Recall Set Size},
    ylabel={Loss in Accuracy},
    xmin=20, xmax=200,
    ymin=-13 , ymax=-2,
    xtick={20, 100, 200},
    ytick={-12,-10,-8,-6,-4,-2},
    legend pos=south east,
    ymajorgrids=true,
    grid style=dashed,
    legend style={nodes={scale=0.4, transform shape}}, 
    legend image post style={mark=*}
]
\addplot[
    color=blue,
    mark=square,
    ]
    coordinates {
    (20, -10.28) (100, -5.07 ) (200,-4.9 )
    };
\addplot[
    color=red,
    mark=square,
    ]
    coordinates {
    (20, -12.89) (100, -7.05 ) (200, -5.39)
    };
\addplot[
    color=green,
    mark=triangle,
    ]
    coordinates {
    (20, -4.95) (100, -2.67 ) (200, -2.76 )
    };
    
\addplot[
    color=purple,
    mark=triangle,
    ]
    coordinates {
    (20, -5.95) (100, -3.46 )(200,-2.94 )
    };
\legend{Baseline, PT, DA, CAPOT }
 \end{axis}}
\end{tikzpicture}
    \centering
    \caption{Average Relative loss in bi-encoder recall accuracy on NQ by recall set size depth on the baseline, Pretrained Alignment (PT), Data Augmentation (DA), and Contrastive Alignment Post Training (CAPOT) on noisy queries.}
    \label{fig:CAPOT-all-nq}
\end{figure}
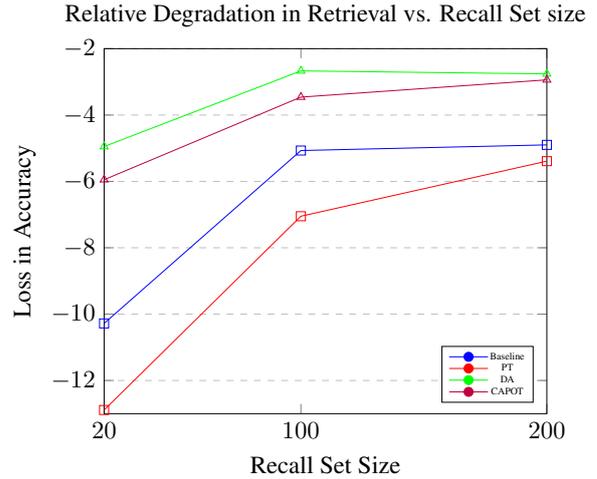
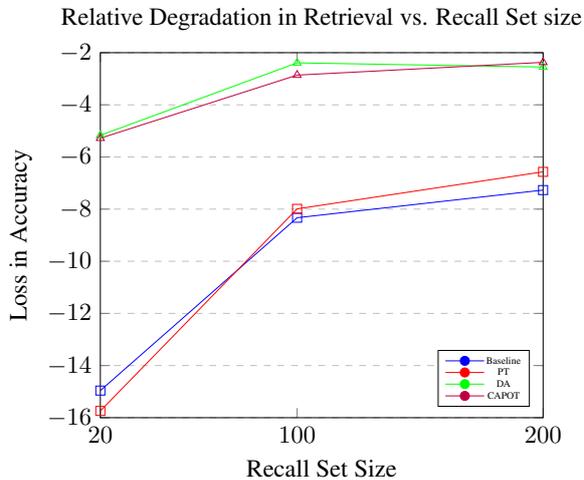
\begin{figure}[!htb]
\begin{tikzpicture}
\scalebox{0.85}{
\begin{axis}[
    title={Relative Degradation in Retrieval  vs. Recall Set size},
    xlabel={Recall Set Size},
    ylabel={Loss in Accuracy},
    xmin=20, xmax=200,
    ymin=-16 , ymax=-2,
    xtick={20, 100, 200},
    ytick={-16,-14,-12,-10,-8,-6,-4,-2},
    legend pos=south east,
    ymajorgrids=true,
    grid style=dashed,
    legend style={nodes={scale=0.4, transform shape}}, 
    legend image post style={mark=*}
]
\addplot[
    color=blue,
    mark=square,
    ]
    coordinates {
    (20, -14.96) (100, -8.33) (200,-7.27 )
    };
\addplot[
    color=red,
    mark=square,
    ]
    coordinates {
    (20, -15.74) (100, -7.99 ) (200, -6.57)
    };
\addplot[
    color=green,
    mark=triangle,
    ]
    coordinates {
    (20, -5.17) (100, -2.39 ) (200, -2.55 )
    };
    
\addplot[
    color=purple,
    mark=triangle,
    ]
    coordinates {
    (20, -5.28) (100, -2.86 )(200,-2.37 )
    };
\legend{Baseline, PT, DA, CAPOT }
 \end{axis}}
\end{tikzpicture}
    \centering
    \caption{Average Relative loss in bi-encoder recall accuracy on NQ by recall set size depth on the baseline, Pretrained Alignment (PT), Data Augmentation (DA), and Contrastive Alignment Post Training (CAPOT) on character-based noisy queries (typos).}
    \label{fig:CAPOT-typo-nq}
\end{figure}

As shown in figures \ref{fig:CAPOT-all-nq} and \ref{fig:CAPOT-typo-nq}, using CAPOT can improve performance on queries with noise, particularly typos. Moreover, the impact of CAPOT is similar to DA without a training set alteration or index regeneration. CAPOT approach takes advantage of training on only the query encoder and fixing the document encoder. Since queries tend to be short, CAPOT uses a max sequence length of 28 tokens to minimize memory usage, allowing scaling to batch sizes of 2048 on GPUs with 16 GBs. This large batch size means training is rapid and effective. A complete alignment run on the NQ dataset takes one hour on a single V100 gpu. At the same time, data augmentation requires 26 hours for training and an additional day for index generation (50 hours overall).\\ 

\begin{table*}[!ht]
    \centering
    \tiny
    \begin{tabular}{|l|l|l|l|l|l|l|l|l|l|l|l|l|}
    \hline
    Dataset & \multicolumn{3}{l}{Regular} &  \multicolumn{3}{l}{DA} & \multicolumn{3}{l}{PT} & \multicolumn{3}{l}{CAPOT}  \\ \hline 
        Depth & 20 & 100 & 200 & 20 & 100 & 200 & 20 & 100 & 200 & 20 & 100 & 200 \\ \hline
        NQ & -10.28\% & -5.07\% & -4.91\% & -4.95\% & -2.67\% & -2.76\% & -12.89\% & -7.05\% & -5.39\% & -5.95\% & -3.46\% & -2.94\% \\ \hline
        TriviaQA & -4.90\% & -2.98\% & -2.24\% & -7.20\% & -4.44\% & -3.57\% & -11.89\% & -6.83\% & -5.34\% & -3.37\% & -1.68\% & -1.17\% \\ \hline
        MSMARCO  & -20.92\% & -33.91\% & -30.46\% & -43.98\% & -28.89\% & -16.73\% & -46.28\% & -36.41\% & -28.69\% & -22.76\% &	-16.73\% & -14.48\% \\ \hline
    \end{tabular}
    \caption{Relative degradation in retrieval accuracy at 20,100,200 on NQ,TriviaQA, and MSMARCO. Retrieval accuracy and relative loss across types of noise for unaltered (Regular),  Data Augmentation (DA),Pre Training Alignment (PT), and Post Training Contrastive Alignment (CAPOT)}
    \label{tab:capot-vs-other-aggregate}
\end{table*}
\begin{table*}[!ht]
    \centering
    \tiny
    \begin{tabular}{|l|l|l|l|l|l|l|l|l|l|l|l|l|}
    \hline
    Dataset & \multicolumn{3}{l}{Regular} &  \multicolumn{3}{l}{DA} & \multicolumn{3}{l}{PT} & \multicolumn{3}{l}{CAPOT}  \\ \hline 
        Depth & 20 & 100 & 200 & 20 & 100 & 200 & 20 & 100 & 200 & 20 & 100 & 200 \\ \hline
        NQ & -14.96\% & -8.33\% & -7.27\% & -5.17\% & -2.39\% & -2.55\% & -15.74\% & -7.99\% & -6.57\% & -5.28\% & -2.86\% & -2.37\% \\ \hline
        TriviaQA & -8.43\% & -4.56\% & -3.39\% & -7.71\% & -4.44\% & -3.64\% & -14.64\% & -8.28\% & -5.47\% & -3.39\% & -1.42\% & -0.87\% \\ \hline
        MSMARCO  & -41.68\% & -33.91\% & -30.46\% & -43.98\% & -33.70\% & -28.69\% & -55.58\% & -45.47\% & -40.95\% & -24.58\% & -18.40\% & -15.95\% \\ \hline
    \end{tabular}
    \caption{Relative degradation in retrieval accuracy at 20,100,200 on NQ,TriviaQA, and MSMARCO. Retrieval accuracy and relative loss across types of character alteration noise (typos) for unaltered (Regular),  Data Augmentation (DA),Pre Training Alignment (PT), and Post Training Contrastive Alignment (CAPOT)}
    \label{tab:capot-vs-other-aggregate-typo}
\end{table*}
Looking at summary metrics in table \ref{tab:capot-vs-other-aggregate} and \ref{tab:capot-vs-other-aggregate-typo}, we can see that the use of pre-training alignment is never optimal and always under-performs un unaltered network. We believe this indicates the importance of introducing noise after training. If introduced prior, the noise will likely be forgotten, and it will hamper the ability to learn a proper, relevant representation. \\

\section{Expanding Contrastive Alignment}
Seeking to explore the impact of variations in alignment query distribution's role, we explore how well CAPOT works with an alignment dataset that differs from the evaluation. To do so, we explore the impact of using the previously discussed Noisy-ORCAS dataset to align noisy queries for TriviaQA. Given the differences in dataset size, we train for the same number of optimization steps with the Noisy-Orcas data as we do with the regular data. \\
\begin{figure}[!htb]
\begin{tikzpicture}
\scalebox{0.9}{
\begin{axis}[
    title={Relative Degradation in Retrieval  vs. Recall Set size},
    xlabel={Recall Set Size},
    ylabel={Loss in Accuracy},
    xmin=20, xmax=200,
    ymin=-5 , ymax=-1,
    xtick={20, 100, 200},
    ytick={-5,-4,-3,-2,-1},
    legend pos=south east,
    ymajorgrids=true,
    grid style=dashed,
    legend style={nodes={scale=0.4, transform shape}}, 
    legend image post style={mark=*}
]
\addplot[
    color=green,
    mark=square,
    ]
    coordinates {
    (20, -4.9) (100, -2.98) (200,-2.24 )
    };
\addplot[
    color=blue,
    mark=square,
    ]
    coordinates {
    (20, -3.37) (100, -1.68) (200,-1.17 )
    };
\addplot[
    color=red,
    mark=square,
    ]
    coordinates {
    (20, -4.55) (100, -2.66 ) (200, -2.94)
    };
\legend{Unaltered, CAPOT, CAPOT-ORCAS }
 \end{axis}}
\end{tikzpicture}
    \centering
    \caption{Average Relative loss in bi-encoder recall accuracy on NQ by recall set size depth on of unaltered,Contrastive Alignment Post Training (CAPOT) and Contrastive Alignment Post Training (CAPOT) ORCAS on TriviaQA.}
    \label{fig:CAPOT-orcas}
\end{figure}
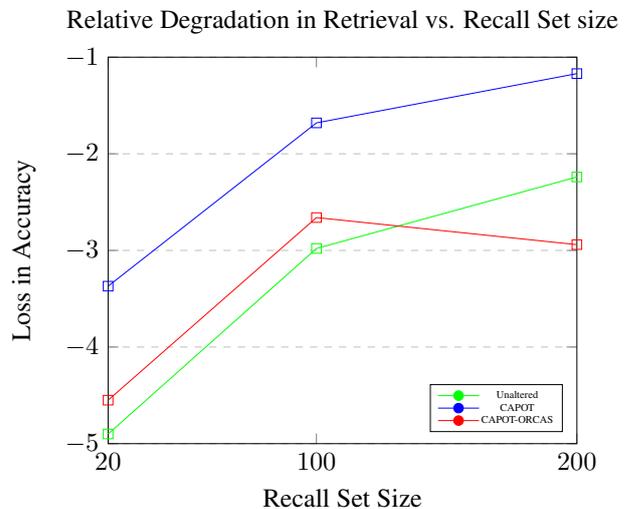 
As shown in figure \ref{fig:CAPOT-orcas}, using an unrelated dataset, ORCAS, provides a close approximation to using the true query distribution, but it does not always outperform the un-altered baseline, indicated by the performance at 200. We believe this is expected as the true query distribution is a factor in how the query vector manifold is optimized.  

%% file: discussion.tex
\begin{figure}[htb!]
    \centering
    \scalebox{0.22}{
    \includegraphics{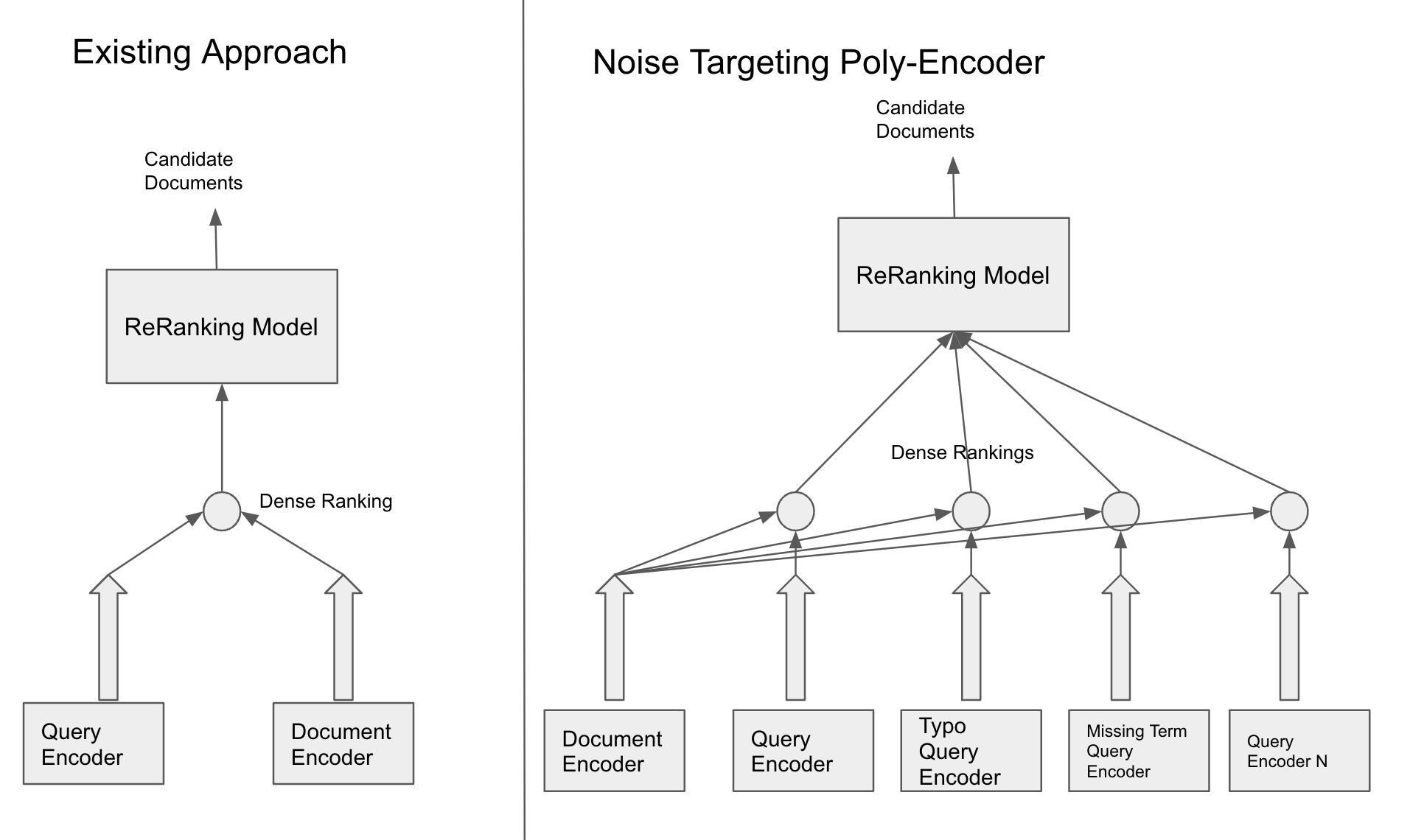}}
    \caption{Proposed poly-encoder architecture using noise-targeted query encoders optimized with CAPOT}
    \label{fig:fig2}
\end{figure}

\textbf{CAPOT and typos} When queries have typos, CAPOT is a computationally efficient method of improving performance as the relative gap between unaltered and aligned is greatest on alterations like character deletion, keyboard character replacement, and random character replacement. We attribute this impact to the relative importance of our alignment dataset's character level alterations. Three out of the 10 methods focus on learning alignments based around minor character shifts, and as a result, the performance optimizes there to the detriment of other forms of noise. CAPOT is much less effective in improving the relevance of minor syntactic shifts such as lemmatization or stemming leading to marginal improvements over the unaltered baselines. We attribute this to the already smaller gap on syntactically altered queries, which on datasets such as TriviaQA have less than 2\% impact.\\
\textbf{CAPOT and Retrieval Set Depth} demonstrates that CAPOT, like DA, sees the highest impact when the recall set size is small. On the NQ the gap between CAPOT and the baseline at 20 is nearly 10\% which narrows to ~3\% at 200.  \\
\textbf{Limitations of contrastive alignment} While effective, contrastive alignment has a non-negligible impact on the retrieval accuracy of unaltered queries. As shown in table \ref{tab:capot-nq-20} on non-noisy queries the use of data augmentation incurs no loss in accuracy yet CAPOT incurs ~2.5\%. This is a fundamental issue because the alignment of embeddings causes minor variations in representations that have actual implications on retrieval accuracy. We believe that the use of larger datasets could such as the web search logs used by the Generic Intent Representation of query vectors \cite{Zhang2019GenericIR} could improve this.\\
\textbf{Poly Encoding} using alignment-based optimization we believe leads to novel retrieval methods which allow for fixed index, constrained optimizations tailored to specific types of noise or deficiencies in retrieval. Novel forms of noise-optimized encoders can be deployed in parallel without additional index generation. Given the prevalence of bi-encoders as candidate set generation tools, CAPOT, unlike Data Augmentation, can generate many targeted query encoder variants which share a document representation. As shown in figure \ref{fig:fig2}, instead of seeking a single query encoder that learns all surface and semantic forms of query representations, alignment approaches can be used to create many encoders which are tuned to various goals. 
\label{app:poly-encoder}

%% file: conclusion.tex
This work studies how to improve bi-encoder performance on noisy queries efficiently. By extending the contrastive triplet loss with an anchoring loss, CAPOT can be used as an approximation for data augmentation without the associated computational overhead. By avoiding retraining and alternating the training corpus, CAPOT can significantly improve recall accuracy with 20 times less computational overhead.  \\ 
In future work, we wish to study how representation alignment approaches can be used with compression approaches such as distillation, pruning, and quantization.

%% file: appendix.tex
\section{Training HyperParameters}
Detailed hyperparameters for bi-encoder training can be found in table \ref{tab:capot-bi-encoder-hy} while the parameters used for post-training alignment can be found in table \ref{tab:capot-post-training-hy}
\begin{table}[htb!]
    \centering
    \scalebox{0.55}{
    \begin{tabular}{|c|c|c|c|c|} \hline
        Dataset & Batch Size & Learning Rate & Train Epochs & Negative Passages \\ \hline
        NQ & 128 & 1e-5 & 40 & 1  \\ \hline
        TrivaQA & 128 & 1e-5 & 40 & 1 \\ \hline
        MSMARCO & 8 & 5e-6 & 3 & 8 \\ \hline
    \end{tabular}}
    \caption{Model Training parameters across tasks. NQ and TriviaQA are trained using 4 V100s, while MSMARCO uses a single V100}
    \label{tab:capot-bi-encoder-hy}
\end{table}
\begin{table}[b!]
    \centering
    \scalebox{0.45}{
        \begin{tabular}{|l|l|l|l|l|l|l|l|l|}
    \hline
        Dataset & Learning Rate  & $\tau_{positive}$ & $\tau_{negative}$ &  $\tau_{anchor}$ &  $\tau_{ranking}$ & Batch Size & Training Time \\ \hline
        NQ &  5.00E-05 & 2 & 0.2 & 2 & 0.7 & 2048 & 1 hour \\ \hline
        TriviaQA &  5.00E-05 & 1 & 0.1 & 2 & 1 & 2048 & 1.5 hours\\ \hline
        MSMARCO & 5.00E-05 & 1 & 0.1 & 2 & 1 & 2048 & 5 hours \\ \hline
    \end{tabular}}
    \caption{CAPOT optimal hyperparameters across datasets. Models were generally trained for 10 epochs, but we find that a single epoch can provide 95\% of the final increase in performance.}
    \label{tab:capot-post-training-hy}
\end{table}
\subsection{Unaltered Bi-encoder With Noise}
\label{sec:full-noise-impact}
Summary results on the impact of noise on bi-encoder retrieval accuracy can be found in \ref{tab:bi-encoder-noise} while full detailed results can be found in tables \ref{tab:nq-r20}, \ref{tab:nq-r100}, \ref{tab:nq-r200}, \ref{tab:msmarco-mrr10}, \ref{tab:msmarco-r20}, \ref{tab:msmarco-r100}, \ref{tab:msmarco-r100}, \ref{tab:msmarco-r200}, \ref{tab:triviaqa-r20}, \ref{tab:triviaqa-r100},\ref{tab:triviaqa-r200}. 
\begin{table*}[htb!]
    \centering
    \scalebox{0.8}{
        \begin{tabular}{|l|l|l|l|l|l|l|l|l|l|}
    \hline
       Dataset & 20 & 20 w/noise & Loss & 100 & 100 w/noise & Loss & 200 & 200 w/noise & Loss \\ \hline
        NQ & 79.73\% & 72.30\% & -9.31\% & 85.98\% & 81.58\% & -5.12\% & 88.25\% & 84.31\% & -4.47\% \\ \hline
        MS Marco & 71.63\% & 58.45\% & -18.40\% & 88.79\% & 58.68\% & -33.91\% & 93.78\% & 80.63\% & -14.02\% \\ \hline
        TriviaQA & 79.40\% & 75.86\% & -4.46\% & 85.01\% & 82.71\% & -2.70\% & 86.66\% & 84.89\% & -2.04\% \\ \hline
    \end{tabular}}
    \caption{Retrieval accuracy for Bi-encoders on unaltered and noisy queries with recall sets of 20,100, and 200 documents. }
    \label{tab:bi-encoder-noise}
\end{table*}
\begin{table*}[htb!]
    \centering
    \scalebox{0.8}{
        \begin{tabular}{|l|l|l|l|l|l|l|l|l|l|}
    \hline
      Dataset & 20 & 20 w/noise & Loss & 100 & 100 w/noise & Impact & 200 & 200 w/noise & Loss \\ \hline
        NQ & 79.73\% & 67.83\% & -14.93\% & 85.98\% & 78.71\% & -8.45\% & 88.25\% & 81.84\% & -7.27\% \\ \hline
        MS Marco & 71.63\% & 41.77\% & -41.69\% & 88.79\% & 70.13\% & -21.02\% & 93.78\% & 82.47\% & -12.06\% \\ \hline
        TriviaQA & 79.40\% & 72.71\% & -8.43\% & 85.01\% & 81.15\% & -4.55\% & 86.66\% & 83.72\% & -3.39\% \\ \hline
    \end{tabular}}
    \caption{Retrieval accuracy for Bi-encoders on unaltered and character-based noisy queries (typos) with recall sets of 20,100, and 200 documents}
    \label{tab:bi-encoder-noise-typos}
\end{table*}
\begin{table*}[htb!]
    \centering
    \tiny
    \scalebox{1}{
        \begin{tabular}{|l|l|l|l|l|l|l|l|l|l|l|}
    \hline
        Noise & Seed 1 & Seed 2 & Seed 3 & Seed 4 & Seed 5 & Median & STD & CI & Upper Bound & Lower Bound \\ \hline
        None & 79.86\% & 79.78\% & 79.47\% & 79.64\% & 79.89\% & 79.73\% & 1.72E-03 & 1.51E-03 & 79.88\% & 79.58\% \\ \hline
        Determiner & 74.46\% & 74.65\% & 74.88\% & 74.54\% & 74.40\% & 74.59\% & 1.86E-03 & 1.63E-03 & 74.75\% & 74.42\% \\ \hline
        Synonym & 68.50\% & 68.23\% & 68.39\% & 68.84\% & 68.37\% & 68.47\% & 2.30E-03 & 2.01E-03 & 68.67\% & 68.26\% \\ \hline
        Lemma & 74.38\% & 74.10\% & 74.46\% & 74.35\% & 74.35\% & 74.33\% & 1.35E-03 & 1.18E-03 & 74.45\% & 74.21\% \\ \hline
        Stem & 74.38\% & 74.10\% & 74.46\% & 73.93\% & 74.35\% & 74.24\% & 2.19E-03 & 1.92E-03 & 74.44\% & 74.05\% \\ \hline
        RCS & 67.06\% & 67.37\% & 67.78\% & 66.62\% & 66.79\% & 67.12\% & 4.65E-03 & 4.08E-03 & 67.53\% & 66.72\% \\ \hline
        KCS & 67.04\% & 67.01\% & 67.40\% & 66.34\% & 67.29\% & 67.01\% & 4.09E-03 & 3.59E-03 & 67.37\% & 66.66\% \\ \hline
        CD & 69.20\% & 69.00\% & 69.97\% & 69.31\% & 69.25\% & 69.35\% & 3.68E-03 & 3.23E-03 & 69.67\% & 69.02\% \\ \hline
        RW & 76.45\% & 76.40\% & 76.48\% & 76.51\% & 76.45\% & 76.46\% & 4.11E-04 & 3.60E-04 & 76.50\% & 76.42\% \\ \hline
        BT & 72.35\% & 72.13\% & 71.91\% & 71.63\% & 72.11\% & 72.03\% & 2.70E-03 & 2.37E-03 & 72.26\% & 71.79\% \\ \hline
        Paraphrase & 72.19\% & 72.19\% & 71.72\% & 71.52\% & 72.38\% & 72.00\% & 3.62E-03 & 3.17E-03 & 72.32\% & 71.68\% \\ \hline
        Average & 72.35\% & 72.27\% & 72.45\% & 72.11\% & 72.33\% & 72.30\% & 1.24E-03 & 1.09E-03 & 72.41\% & 72.19\% \\ \hline
        Typos & 67.77\% & 67.79\% & 68.38\% & 67.42\% & 67.77\% & 67.83\% & 4.14E-03 & 3.63E-03 & 68.19\% & 67.47\% \\ \hline
    \end{tabular}}
    \caption{Baseline retrieval accuracy at recall set size 20 Performance on NQ with 5 random seeds with and without noisy queries.}
    \label{tab:nq-r20}
\end{table*}

\begin{table*}[htb!]
    \centering
    \tiny
    \scalebox{1}{
        \begin{tabular}{|l|l|l|l|l|l|l|l|l|l|l|}
    \hline
       Noise & Seed 1 & Seed 2 & Seed 3 & Seed 4 & Seed 5 & Median & STD & CI & Upper Bound & Lower Bound \\ \hline
        None & 85.98\% & 86.01\% & 86.15\% & 85.93\% & 85.82\% & 85.98\% & 1.21E-03 & 1.06E-03 & 86.08\% & 85.87\% \\ \hline
        Determiner & 83.49\% & 83.35\% & 83.41\% & 83.57\% & 83.43\% & 83.45\% & 8.45E-04 & 7.40E-04 & 83.53\% & 83.38\% \\ \hline
        Synonym & 79.61\% & 79.36\% & 79.78\% & 79.58\% & 79.75\% & 79.62\% & 1.66E-03 & 1.45E-03 & 79.76\% & 79.47\% \\ \hline
        Lemma & 82.85\% & 83.21\% & 82.94\% & 82.71\% & 82.96\% & 82.94\% & 1.83E-03 & 1.60E-03 & 83.10\% & 82.78\% \\ \hline
        Stem & 82.88\% & 83.21\% & 82.94\% & 82.71\% & 82.96\% & 82.94\% & 1.80E-03 & 1.58E-03 & 83.10\% & 82.78\% \\ \hline
        RCS & 78.50\% & 79.00\% & 79.11\% & 77.92\% & 78.78\% & 78.66\% & 4.76E-03 & 4.17E-03 & 79.08\% & 78.25\% \\ \hline
        KCS & 78.01\% & 77.92\% & 78.42\% & 77.73\% & 78.03\% & 78.02\% & 2.53E-03 & 2.22E-03 & 78.24\% & 77.80\% \\ \hline
        CD & 79.75\% & 79.36\% & 79.53\% & 79.00\% & 79.64\% & 79.46\% & 2.92E-03 & 2.56E-03 & 79.71\% & 79.20\% \\ \hline
        RW & 85.07\% & 85.07\% & 84.85\% & 84.82\% & 85.04\% & 84.97\% & 1.25E-03 & 1.09E-03 & 85.08\% & 84.86\% \\ \hline
        BT & 80.14\% & 79.92\% & 80.03\% & 79.47\% & 80.11\% & 79.93\% & 2.71E-03 & 2.38E-03 & 80.17\% & 79.70\% \\ \hline
        Paraphrase & 81.44\% & 81.44\% & 81.19\% & 81.44\% & 81.30\% & 81.36\% & 1.13E-03 & 9.92E-04 & 81.46\% & 81.26\% \\ \hline
        Average & 81.61\% & 81.62\% & 81.67\% & 81.35\% & 81.62\% & 81.58\% & 1.25E-03 & 1.10E-03 & 81.69\% & 81.47\% \\ \hline
        Typos & 78.75\% & 78.76\% & 79.02\% & 78.22\% & 78.82\% & 78.71\% & 0.34\% & 0.30\% & 79.01\% & 78.42\% \\ \hline
    \end{tabular}}
    \caption{Baseline retrieval accuracy at recall set size 100 Performance on NQ with 5 random seeds with and without noisy queries.}
    \label{tab:nq-r100}
\end{table*}

\begin{table*}[htb!]
    \centering
    \tiny
    \scalebox{1}{
        \begin{tabular}{|l|l|l|l|l|l|l|l|l|l|l|}
    \hline
        Noise & Seed 1 & Seed 2 & Seed 3 & Seed 4 & Seed 5 & Median & STD & CI & Upper Bound & Lower Bound \\ \hline
        None & 88.42\% & 87.89\% & 88.23\% & 88.28\% & 88.42\% & 88.25\% & 2.16E-03 & 1.89E-03 & 88.44\% & 88.06\% \\ \hline
        Determiner & 86.09\% & 85.82\% & 86.04\% & 85.87\% & 86.12\% & 85.99\% & 1.36E-03 & 1.19E-03 & 86.11\% & 85.87\% \\ \hline
        Synonym & 82.85\% & 82.85\% & 82.66\% & 82.80\% & 82.96\% & 82.83\% & 1.11E-03 & 9.71E-04 & 82.92\% & 82.73\% \\ \hline
        Lemma & 86.04\% & 85.37\% & 82.94\% & 85.37\% & 85.93\% & 85.13\% & 1.26E-02 & 1.11E-02 & 86.24\% & 84.02\% \\ \hline
        Stem & 86.01\% & 85.37\% & 85.90\% & 85.37\% & 85.93\% & 85.72\% & 3.16E-03 & 2.77E-03 & 85.99\% & 85.44\% \\ \hline
        RCS & 81.72\% & 82.08\% & 82.49\% & 81.86\% & 81.63\% & 81.96\% & 3.44E-03 & 3.02E-03 & 82.26\% & 81.65\% \\ \hline
        KCS & 81.69\% & 81.47\% & 81.72\% & 80.97\% & 81.75\% & 81.52\% & 3.26E-03 & 2.85E-03 & 81.80\% & 81.23\% \\ \hline
        CD & 79.75\% & 82.60\% & 82.83\% & 82.44\% & 82.58\% & 82.04\% & 1.29E-02 & 1.13E-02 & 83.17\% & 80.91\% \\ \hline
        RW & 87.34\% & 87.40\% & 87.26\% & 87.34\% & 87.37\% & 87.34\% & 5.18E-04 & 4.54E-04 & 87.39\% & 87.30\% \\ \hline
        BT & 82.52\% & 82.19\% & 82.94\% & 82.47\% & 82.60\% & 82.54\% & 2.70E-03 & 2.36E-03 & 82.78\% & 82.31\% \\ \hline
        Paraphrase & 83.93\% & 83.93\% & 83.99\% & 84.27\% & 84.32\% & 84.09\% & 1.90E-03 & 1.66E-03 & 84.25\% & 83.92\% \\ \hline
        Average & 84.22\% & 84.27\% & 84.27\% & 84.28\% & 84.51\% & 84.31\% & 1.15E-03 & 1.01E-03 & 84.41\% & 84.21\% \\ \hline
        Typos & 81.05\% & 82.05\% & 82.35\% & 81.75\% & 81.99\% & 81.84\% & 6.52E-03 & 5.72E-03 & 82.41\% & 81.27\% \\ \hline
    \end{tabular}}
    \caption{Baseline retrieval accuracy at recall set size 200 Performance on NQ with 5 random seeds with and without noisy queries.}
    \label{tab:nq-r200}
\end{table*}
\begin{table*}[htb!]
    \centering
    \tiny
    \scalebox{1}{
        \begin{tabular}{|l|l|l|l|l|l|l|l|l|l|l|}
    \hline
    Noise & Seed 1 & Seed 2 & Seed 3 & Seed 4 & Seed 5 & Median & STD & CI & Upper Bound & Lower Bound \\ \hline
        None & 79.37\% & 79.56\% & 79.58\% & 79.32\% & 79.17\% & 79.40\% & 1.71E-03 & 1.50E-03 & 79.55\% & 79.25\% \\ \hline
        Determiner & 77.40\% & 77.44\% & 77.39\% & 77.54\% & 77.02\% & 77.36\% & 1.99E-03 & 1.74E-03 & 77.53\% & 77.18\% \\ \hline
        Synonym & 74.93\% & 75.18\% & 75.06\% & 75.21\% & 75.15\% & 75.11\% & 1.11E-03 & 9.76E-04 & 75.20\% & 75.01\% \\ \hline
        Lemma & 79.11\% & 79.11\% & 79.15\% & 78.95\% & 79.02\% & 79.06\% & 8.91E-04 & 7.81E-04 & 79.14\% & 78.98\% \\ \hline
        Stem & 78.11\% & 78.35\% & 78.41\% & 77.98\% & 78.18\% & 78.21\% & 1.77E-03 & 1.55E-03 & 78.36\% & 78.05\% \\ \hline
        RCS & 72.41\% & 72.94\% & 72.67\% & 72.60\% & 72.60\% & 72.64\% & 1.92E-03 & 1.68E-03 & 72.81\% & 72.48\% \\ \hline
        KCS & 72.24\% & 72.39\% & 72.52\% & 72.34\% & 72.21\% & 72.34\% & 1.23E-03 & 1.08E-03 & 72.45\% & 72.23\% \\ \hline
        CD & 73.31\% & 73.16\% & 73.09\% & 73.10\% & 73.06\% & 73.14\% & 9.77E-04 & 8.56E-04 & 73.23\% & 73.06\% \\ \hline
        RW & 78.16\% & 78.18\% & 78.16\% & 77.80\% & 77.96\% & 78.05\% & 1.69E-03 & 1.48E-03 & 78.20\% & 77.90\% \\ \hline
        BT & 73.76\% & 73.93\% & 73.77\% & 75.08\% & 73.69\% & 74.05\% & 5.85E-03 & 5.13E-03 & 74.56\% & 73.53\% \\ \hline
        Paraphrase & 75.04\% & 75.19\% & 75.36\% & 75.08\% & 74.84\% & 75.10\% & 1.89E-03 & 1.66E-03 & 75.27\% & 74.94\% \\ \hline
        Average & 75.80\% & 75.95\% & 75.92\% & 75.91\% & 75.72\% & 75.86\% & 9.75E-04 & 8.54E-04 & 75.95\% & 75.78\% \\ \hline
        Typos & 72.65\% & 72.83\% & 72.76\% & 72.68\% & 72.62\% & 72.71\% & 1.38E-03 & 1.21E-03 & 72.83\% & 72.59\% \\ \hline
    \end{tabular}}
    \caption{Baseline retrieval accuracy at recall set size 20 Performance on TrivaQA  with 5 random seeds with and without noisy queries.}
    \label{tab:triviaqa-r20}
\end{table*}

\begin{table*}[htb!]
    \centering
    \tiny
    \scalebox{1}{
        \begin{tabular}{|l|l|l|l|l|l|l|l|l|l|l|}
    \hline
     Noise & Seed 1 & Seed 2 & Seed 3 & Seed 4 & Seed 5 & Median & STD & CI & Upper Bound & Lower Bound \\ \hline
        None & 85.03\% & 84.95\% & 85.10\% & 85.10\% & 84.88\% & 85.01\% & 9.41E-04 & 8.25E-04 & 85.09\% & 84.93\% \\ \hline
        Determiner & 83.78\% & 83.77\% & 84.02\% & 83.90\% & 83.70\% & 83.83\% & 1.26E-03 & 1.11E-03 & 83.95\% & 83.72\% \\ \hline
        Synonym & 82.37\% & 82.41\% & 82.52\% & 82.37\% & 82.27\% & 82.39\% & 8.87E-04 & 7.78E-04 & 82.47\% & 82.31\% \\ \hline
        Lemma & 84.81\% & 84.82\% & 84.82\% & 84.92\% & 84.68\% & 84.81\% & 8.51E-04 & 7.46E-04 & 84.89\% & 84.74\% \\ \hline
        Stem & 84.40\% & 84.32\% & 84.45\% & 84.42\% & 84.20\% & 84.36\% & 1.00E-03 & 8.78E-04 & 84.45\% & 84.27\% \\ \hline
        RCS & 80.92\% & 81.00\% & 81.00\% & 80.92\% & 80.92\% & 80.95\% & 4.13E-04 & 3.62E-04 & 80.99\% & 80.92\% \\ \hline
        KCS & 81.12\% & 81.03\% & 81.39\% & 81.07\% & 81.19\% & 81.16\% & 1.44E-03 & 1.26E-03 & 81.29\% & 81.03\% \\ \hline
        CD & 81.34\% & 81.26\% & 81.38\% & 81.35\% & 81.30\% & 81.33\% & 4.83E-04 & 4.23E-04 & 81.37\% & 81.28\% \\ \hline
        RW & 84.33\% & 84.27\% & 84.36\% & 77.80\% & 84.20\% & 82.99\% & 2.91E-02 & 2.55E-02 & 85.54\% & 80.45\% \\ \hline
        BT & 80.33\% & 80.08\% & 80.20\% & 82.26\% & 80.12\% & 80.60\% & 9.33E-03 & 8.18E-03 & 81.42\% & 79.78\% \\ \hline
        Paraphrase & 82.46\% & 82.50\% & 82.33\% & 82.50\% & 82.21\% & 82.40\% & 1.28E-03 & 1.12E-03 & 82.51\% & 82.29\% \\ \hline
        Average & 82.81\% & 82.77\% & 82.87\% & 82.42\% & 82.70\% & 82.71\% & 1.76E-03 & 1.54E-03 & 82.87\% & 82.56\% \\ \hline
        Typos & 81.13\% & 81.10\% & 81.26\% & 81.11\% & 81.14\% & 81.15\% & 0.08\% & 0.07\% & 81.21\% & 81.08\% \\ \hline   
    \end{tabular}}
    \caption{Baseline retrieval accuracy at recall set size 100 Performance on TrivaQA  with 5 random seeds with and without noisy queries.}
    \label{tab:triviaqa-r100}
\end{table*}

\begin{table*}[htb!]
    \centering
    \tiny
    \scalebox{1}{
        \begin{tabular}{|l|l|l|l|l|l|l|l|l|l|l|}
    \hline
      Noise & Seed 1 & Seed 2 & Seed 3 & Seed 4 & Seed 5 & Median & STD & CI & Upper Bound & Lower Bound \\ \hline
        None & 86.87\% & 86.53\% & 86.61\% & 86.63\% & 86.67\% & 86.66\% & 1.29E-03 & 1.13E-03 & 86.77\% & 86.55\% \\ \hline
        Determiner & 85.73\% & 85.66\% & 85.62\% & 85.65\% & 85.73\% & 85.68\% & 5.12E-04 & 4.48E-04 & 85.73\% & 85.64\% \\ \hline
        Synonym & 84.58\% & 84.46\% & 84.46\% & 84.67\% & 84.58\% & 84.55\% & 8.99E-04 & 7.88E-04 & 84.63\% & 84.47\% \\ \hline
        Lemma & 86.72\% & 86.35\% & 86.45\% & 86.44\% & 86.54\% & 86.50\% & 1.41E-03 & 1.23E-03 & 86.62\% & 86.38\% \\ \hline
        Stem & 86.38\% & 86.04\% & 86.09\% & 86.15\% & 86.24\% & 86.18\% & 1.33E-03 & 1.17E-03 & 86.30\% & 86.06\% \\ \hline
        RCS & 83.64\% & 83.43\% & 83.48\% & 83.55\% & 83.55\% & 83.53\% & 7.77E-04 & 6.81E-04 & 83.60\% & 83.46\% \\ \hline
        KCS & 83.92\% & 83.72\% & 83.89\% & 83.66\% & 83.80\% & 83.80\% & 1.13E-03 & 9.91E-04 & 83.90\% & 83.70\% \\ \hline
        CD & 83.81\% & 83.86\% & 83.77\% & 83.85\% & 83.93\% & 83.84\% & 6.01E-04 & 5.27E-04 & 83.90\% & 83.79\% \\ \hline
        RW & 86.05\% & 86.32\% & 85.98\% & 86.09\% & 85.97\% & 86.08\% & 1.40E-03 & 1.23E-03 & 86.20\% & 85.96\% \\ \hline
        BT & 82.39\% & 82.07\% & 82.19\% & 84.35\% & 82.22\% & 82.64\% & 9.58E-03 & 8.39E-03 & 83.48\% & 81.81\% \\ \hline
        Paraphrase & 84.42\% & 84.35\% & 84.32\% & 84.35\% & 84.39\% & 84.36\% & 3.90E-04 & 3.42E-04 & 84.40\% & 84.33\% \\ \hline
        Average & 84.96\% & 84.80\% & 84.81\% & 85.03\% & 84.87\% & 84.89\% & 1.01E-03 & 8.83E-04 & 84.98\% & 84.81\% \\ \hline
        Typos & 83.79\% & 83.67\% & 83.71\% & 83.69\% & 83.76\% & 83.72\% & 8.36E-04 & 7.33E-04 & 83.80\% & 83.65\% \\ \hline
    \end{tabular}}
    \caption{Baseline retrieval accuracy at recall set size 200 Performance on TrivaQA with 5 random seeds with and without noisy queries.}
    \label{tab:triviaqa-r200}
\end{table*}

\begin{table*}[htb!]
    \centering
    \tiny
    \scalebox{1}{
        \begin{tabular}{|l|l|l|l|l|l|l|l|l|l|l|}
    \hline
    Noise & Seed 1 & Seed 2 & Seed 3 & Seed 3 & Seed 5 & Median & STD & CI & Upper Bound & Lower Bound \\ \hline
        None & 32.43\% & 32.41\% & 32.31\% & 32.81\% & 32.01\% & 32.39\% & 2.87E-03 & 2.51E-03 & 3.26E-01 & 3.21E-01 \\ \hline
        Determiner & 25.86\% & 26.01\% & 25.55\% & 26.12\% & 25.71\% & 25.85\% & 2.30E-03 & 2.02E-03 & 2.61E-01 & 2.56E-01 \\ \hline
        Synonym & 20.75\% & 20.52\% & 20.36\% & 20.94\% & 20.58\% & 20.63\% & 2.23E-03 & 1.96E-03 & 2.08E-01 & 2.04E-01 \\ \hline
        Lemma & 31.55\% & 31.54\% & 31.50\% & 31.84\% & 31.10\% & 31.50\% & 2.65E-03 & 2.32E-03 & 3.17E-01 & 3.13E-01 \\ \hline
        Stem & 26.88\% & 27.01\% & 26.65\% & 27.23\% & 26.59\% & 26.87\% & 2.63E-03 & 2.31E-03 & 2.71E-01 & 2.66E-01 \\ \hline
        RCS & 16.87\% & 16.70\% & 16.46\% & 17.36\% & 16.89\% & 16.85\% & 3.29E-03 & 2.88E-03 & 1.71E-01 & 1.66E-01 \\ \hline
        KCS & 15.93\% & 16.06\% & 15.63\% & 16.48\% & 15.88\% & 16.00\% & 3.14E-03 & 2.75E-03 & 1.63E-01 & 1.57E-01 \\ \hline
        CD & 18.17\% & 17.91\% & 17.63\% & 18.46\% & 17.87\% & 18.00\% & 3.16E-03 & 2.77E-03 & 1.83E-01 & 1.77E-01 \\ \hline
        RW & 31.37\% & 31.30\% & 30.99\% & 31.36\% & 30.96\% & 31.19\% & 2.05E-03 & 1.80E-03 & 3.14E-01 & 3.10E-01 \\ \hline
        BT & 26.95\% & 26.57\% & 26.62\% & 27.10\% & 26.54\% & 26.76\% & 2.55E-03 & 2.23E-03 & 2.70E-01 & 2.65E-01 \\ \hline
        Paraphrase & 26.69\% & 26.51\% & 26.38\% & 26.75\% & 26.50\% & 26.56\% & 1.55E-03 & 1.35E-03 & 2.67E-01 & 2.64E-01 \\ \hline
        Average & 24.86\% & 24.78\% & 24.55\% & 25.13\% & 24.60\% & 24.78\% & 2.32E-03 & 2.03E-03 & 2.50E-01 & 2.46E-01 \\ \hline
        Typos & 16.99\% & 16.89\% & 16.57\% & 17.43\% & 16.88\% & 16.95\% & 3.20E-03 & 2.80E-03 & 1.72E-01 & 1.67E-01 \\ \hline 
    \end{tabular}}
    \caption{Dense Model MRR@10 Performance on MSMARCO with 5 random seeds with and without noisy queries.}
    \label{tab:msmarco-mrr10}
\end{table*}
\begin{table*}[htb!]
    \centering
    \tiny
    \scalebox{1}{
        \begin{tabular}{|l|l|l|l|l|l|l|l|l|l|l|}
    \hline
     Noise & Seed 1 & Seed 2 & Seed 3 & Seed 3 & Seed 5 & Median & STD & CI & Upper Bound & Lower Bound \\ \hline
        None & 71.36\% & 72.03\% & 71.65\% & 71.83\% & 71.29\% & 71.63\% & 3.14E-03 & 2.75E-03 & 71.91\% & 71.36\% \\ \hline
        Determiner & 59.64\% & 60.20\% & 59.91\% & 60.10\% & 59.46\% & 59.86\% & 3.11E-03 & 2.73E-03 & 60.14\% & 59.59\% \\ \hline
        Synonym & 49.96\% & 50.09\% & 49.63\% & 50.46\% & 50.37\% & 50.10\% & 3.34E-03 & 2.93E-03 & 50.39\% & 49.81\% \\ \hline
        Lemma & 69.84\% & 70.24\% & 92.71\% & 70.09\% & 69.99\% & 74.57\% & 1.01E-01 & 8.89E-02 & 83.46\% & 65.69\% \\ \hline
        Stem & 61.91\% & 61.88\% & 61.70\% & 61.81\% & 79.23\% & 65.30\% & 7.78E-02 & 6.82E-02 & 72.13\% & 58.48\% \\ \hline
        RCS & 41.33\% & 41.63\% & 40.87\% & 41.86\% & 40.80\% & 41.30\% & 4.63E-03 & 4.06E-03 & 41.71\% & 40.89\% \\ \hline
        KCS & 39.66\% & 40.01\% & 39.18\% & 40.04\% & 39.20\% & 39.62\% & 4.20E-03 & 3.68E-03 & 39.99\% & 39.25\% \\ \hline
        CD & 43.81\% & 44.77\% & 44.03\% & 44.80\% & 44.57\% & 44.40\% & 4.51E-03 & 3.95E-03 & 44.79\% & 44.00\% \\ \hline
        RW & 69.84\% & 70.06\% & 69.63\% & 69.96\% & 69.91\% & 61.23\% & 2.03E-03 & 1.78E-03 & 61.40\% & 61.05\% \\ \hline
        BT & 61.23\% & 61.40\% & 60.99\% & 61.45\% & 61.06\% & 65.05\% & 8.19E-02 & 7.18E-02 & 72.23\% & 57.87\% \\ \hline
        Paraphrase & 61.28\% & 61.83\% & 79.70\% & 61.30\% & 61.15\% & 65.05\% & 8.19E-02 & 7.18E-02 & 72.23\% & 57.87\% \\ \hline
        Average & 57.26\% & 57.65\% & 60.91\% & 57.61\% & 58.82\% & 58.45\% & 1.50E-02 & 1.31E-02 & 59.76\% & 57.14\% \\ \hline
        Typos & 41.60\% & 42.14\% & 41.36\% & 42.23\% & 41.52\% & 41.77\% & 0.44\% & 0.39\% & 42.16\% & 41.38\% \\ \hline
    \end{tabular}}
    \caption{Baseline retrieval accuracy at recall set size 20 Performance on MSMARCO with 5 random seeds with and without noisy queries.}
    \label{tab:msmarco-r20}
\end{table*}

\begin{table*}[htb!]
    \centering
    \tiny
    \scalebox{1}{
        \begin{tabular}{|l|l|l|l|l|l|l|l|l|l|l|}
    \hline
    Noise & Seed 1 & Seed 2 & Seed 3 & Seed 3 & Seed 5 & Median & STD & CI & Upper Bound & Lower Bound \\ \hline
        None & 88.40\% & 89.10\% & 88.67\% & 89.05\% & 88.75\% & 88.79\% & 2.90E-03 & 2.54E-03 & 89.05\% & 88.54\% \\ \hline
        Determiner & 77.36\% & 77.69\% & 77.55\% & 77.89\% & 77.79\% & 77.66\% & 2.08E-03 & 1.83E-03 & 77.84\% & 77.48\% \\ \hline
        Synonym & 67.88\% & 68.67\% & 67.52\% & 68.24\% & 68.14\% & 68.09\% & 4.26E-03 & 3.73E-03 & 68.46\% & 67.72\% \\ \hline
        Lemma & 87.08\% & 87.81\% & 87.39\% & 87.85\% & 87.55\% & 87.54\% & 3.18E-03 & 2.79E-03 & 87.81\% & 87.26\% \\ \hline
        Stem & 78.91\% & 79.76\% & 79.11\% & 79.50\% & 79.23\% & 79.30\% & 3.32E-03 & 2.91E-03 & 79.59\% & 79.01\% \\ \hline
        RCS & 57.46\% & 58.38\% & 57.32\% & 58.90\% & 57.92\% & 58.00\% & 6.53E-03 & 5.72E-03 & 58.57\% & 57.43\% \\ \hline
        KCS & 55.72\% & 56.07\% & 55.70\% & 56.79\% & 56.15\% & 56.09\% & 5.73E-03 & 5.02E-03 & 56.59\% & 55.58\% \\ \hline
        CD & 61.49\% & 61.92\% & 61.78\% & 62.95\% & 61.69\% & 61.97\% & 3.56E-03 & 3.12E-03 & 62.28\% & 61.65\% \\ \hline
        RW & 87.31\% & 87.84\% & 87.36\% & 87.77\% & 87.58\% & 87.57\% & 2.57E-03 & 2.25E-03 & 87.80\% & 87.35\% \\ \hline
        BT & 78.35\% & 78.91\% & 78.12\% & 78.94\% & 78.68\% & 78.60\% & 3.56E-03 & 3.12E-03 & 78.91\% & 78.29\% \\ \hline
        Paraphrase & 79.44\% & 79.74\% & 79.70\% & 80.04\% & 79.41\% & 79.67\% & 2.57E-03 & 2.25E-03 & 79.89\% & 79.44\% \\ \hline
        Average & 74.49\% & 75.08\% & 74.57\% & 75.27\% & 74.81\% & 74.84\% & 3.31E-03 & 2.90E-03 & 75.13\% & 74.55\% \\ \hline
        Typos & 58.22\% & 58.79\% & 58.27\% & 59.55\% & 58.59\% & 58.68\% & 5.27E-03 & 4.62E-03 & 59.14\% & 58.22\% \\ \hline
    \end{tabular}}
    \caption{Baseline retrieval accuracy at recall set size 100 Performance on MSMARCO with 5 random seeds with and without noisy queries.}
    \label{tab:msmarco-r100}
\end{table*}

\begin{table*}[htb!]
    \centering
    \tiny
    \scalebox{1}{
        \begin{tabular}{|l|l|l|l|l|l|l|l|l|l|l|}
    \hline
    Noise & Seed 1 & Seed 2 & Seed 3 & Seed 3 & Seed 5 & Median & STD & CI & Upper Bound & Lower Bound \\ \hline
        None & 93.93\% & 93.95\% & 93.34\% & 93.90\% & 93.80\% & 93.78\% & 2.55E-03 & 2.24E-03 & 94.01\% & 93.56\% \\ \hline
        Determiner & 83.38\% & 83.90\% & 83.44\% & 83.67\% & 83.44\% & 83.56\% & 2.16E-03 & 1.89E-03 & 83.75\% & 83.38\% \\ \hline
        Synonym & 74.27\% & 75.03\% & 74.01\% & 74.76\% & 74.63\% & 74.54\% & 4.02E-03 & 3.52E-03 & 74.89\% & 74.19\% \\ \hline
        Lemma & 92.68\% & 92.71\% & 92.35\% & 92.65\% & 92.48\% & 92.57\% & 1.54E-03 & 1.35E-03 & 92.71\% & 92.44\% \\ \hline
        Stem & 84.74\% & 84.83\% & 84.30\% & 84.91\% & 84.70\% & 84.70\% & 2.37E-03 & 2.08E-03 & 84.90\% & 84.49\% \\ \hline
        RCS & 63.95\% & 64.66\% & 63.81\% & 65.32\% & 64.36\% & 64.42\% & 6.02E-03 & 5.28E-03 & 64.95\% & 63.89\% \\ \hline
        KCS & 62.05\% & 62.97\% & 62.61\% & 63.84\% & 62.91\% & 62.87\% & 6.51E-03 & 5.70E-03 & 63.44\% & 62.30\% \\ \hline
        CD & 68.38\% & 68.47\% & 67.66\% & 68.88\% & 68.38\% & 68.36\% & 4.38E-03 & 3.84E-03 & 68.74\% & 67.97\% \\ \hline
        RW & 92.58\% & 92.78\% & 92.56\% & 92.71\% & 92.52\% & 92.63\% & 1.08E-03 & 9.51E-04 & 92.73\% & 92.54\% \\ \hline
        BT & 84.21\% & 84.36\% & 83.72\% & 83.97\% & 84.14\% & 84.08\% & 2.43E-03 & 2.13E-03 & 84.29\% & 83.87\% \\ \hline
        Paraphrase & 85.60\% & 85.79\% & 85.70\% & 86.39\% & 85.60\% & 85.82\% & 3.30E-03 & 2.89E-03 & 86.11\% & 85.53\% \\ \hline
        Average & 80.52\% & 80.86\% & 80.32\% & 81.00\% & 80.63\% & 80.67\% & 2.69E-03 & 2.36E-03 & 80.90\% & 80.43\% \\ \hline
        Typos & 64.79\% & 65.36\% & 64.69\% & 66.01\% & 65.21\% & 65.22\% & 5.64E-03 & 4.94E-03 & 65.71\% & 64.72\% \\ \hline 
    \end{tabular}}
    \caption{Baseline retrieval accuracy at recall set size 200 Performance on MSMARCO with five random seeds with and without noisy queries.}
    \label{tab:msmarco-r200}
\end{table*}
\subsection{CAPOT full Results}
\label{app:capot}
\input{appendix-capot}
\subsection{CAPOT and ORCAS}
\label{app:capot-orcas}
\begin{table*}[!ht]
    \centering
    \begin{tabular}{|l|l|l|l|l|l|l|}
    \hline
        Noise & Baseline & Loss & CAPOT & Loss & CAPOT-ORCAS & Loss \\ \hline
        None & 79.40\% & 0.00\% & 78.53\% & -1.10\% & 76.73\% & -3.36\% \\ \hline
        Determiner & 77.36\% & -2.57\% & 77.76\% & -2.07\% & 76.04\% & -4.24\% \\ \hline
        Synonym & 75.11\% & -5.41\% & 75.97\% & -4.33\% & 81.23\% & 2.31\% \\ \hline
        Lemma & 79.06\% & -0.43\% & 78.69\% & -0.90\% & 76.95\% & -3.09\% \\ \hline
        Stem & 78.21\% & -1.50\% & 78.24\% & -1.46\% & 76.68\% & -3.42\% \\ \hline
        RCS & 72.64\% & -8.51\% & 76.74\% & -3.35\% & 75.83\% & -4.49\% \\ \hline
        KCS & 72.34\% & -8.89\% & 76.90\% & -3.15\% & 75.62\% & -4.76\% \\ \hline
        CD & 73.14\% & -7.88\% & 76.47\% & -3.69\% & 75.28\% & -5.19\% \\ \hline
        RW & 78.05\% & -1.70\% & 78.74\% & -0.83\% & 76.81\% & -3.27\% \\ \hline
        BT & 74.05\% & -6.74\% & 73.56\% & -7.35\% & 71.48\% & -9.97\% \\ \hline
        Paraphrase & 75.10\% & -5.41\% & 74.19\% & -6.56\% & 71.95\% & -9.38\% \\ \hline
        Average & 75.51\% & -4.90\% & 76.73\% & -3.37\% & 75.79\% & -4.55\% \\ \hline
        Typos & 72.71\% & -8.43\% & 76.71\% & -3.39\% & 75.58\% & -4.82\% \\ \hline
    \end{tabular}
    \caption{Retrieval accuracy and relative loss across types of noise for unaltered (Regular),  Post Training Contrastive Alignment (CAPOT), and Post Training Contrastive Alignment (CAPOT) using Noisy-ORCAS on TriviaQA dataset with the recall set the size of 20}
    \label{tab:capot-orcasvsnot-20}
\end{table*}
\begin{table*}[!ht]
    \centering
    \begin{tabular}{|l|l|l|l|l|l|l|}
    \hline
        Noise & Baseline & Loss & CAPOT & Loss & CAPOT-ORCAS & Loss \\ \hline
        None & 85.01\% & 0.00\% & 84.85\% & -0.19\% & 85.63\% & 0.73\% \\ \hline
        Determiner & 83.83\% & -1.38\% & 84.25\% & -0.90\% & 83.27\% & -2.05\% \\ \hline
        Synonym & 82.39\% & -3.08\% & 83.12\% & -2.23\% & 81.23\% & -4.44\% \\ \hline
        Lemma & 84.81\% & -0.23\% & 84.93\% & -0.10\% & 83.62\% & -1.63\% \\ \hline
        Stem & 84.36\% & -0.77\% & 84.82\% & -0.22\% & 83.58\% & -1.69\% \\ \hline
        RCS & 80.95\% & -4.77\% & 83.72\% & -1.52\% & 85.04\% & 0.04\% \\ \hline
        KCS & 81.16\% & -4.53\% & 83.95\% & -1.25\% & 85.39\% & 0.45\% \\ \hline
        CD & 81.33\% & -4.33\% & 83.75\% & -1.48\% & 82.66\% & -2.77\% \\ \hline
        RW & 82.99\% & -2.37\% & 84.86\% & -0.18\% & 83.56\% & -1.71\% \\ \hline
        BT & 80.60\% & -5.19\% & 80.16\% & -5.70\% & 78.77\% & -7.34\% \\ \hline
        Paraphrase & 82.40\% & -3.07\% & 82.29\% & -3.20\% & 80.33\% & -5.50\% \\ \hline
        Average & 82.48\% & -2.97\% & 83.58\% & -1.68\% & 82.74\% & -2.66\% \\ \hline
        Typos & 81.15\% & -4.54\% & 83.81\% & -1.42\% & 84.36\% & -0.76\% \\ \hline
    \end{tabular}
    \caption{Retrieval accuracy and relative loss across types of noise for unaltered (Regular),  Post Training Contrastive Alignment (CAPOT), and Post Training Contrastive Alignment (CAPOT) using Noisy-ORCAS on TriviaQA dataset with the recall set the size of 100}
    \label{tab:capot-orcasvsnot-100}
\end{table*}
\begin{table*}[!ht]
    \centering
    \begin{tabular}{|l|l|l|l|l|l|l|}
    \hline
        Noise & Baseline & Loss & CAPOT & Loss & CAPOT-ORCAS & Loss \\ \hline
        None & 86.66\% & 0.00\% & 86.60\% & -0.07\% & 83.52\% & -3.62\% \\ \hline
        Determiner & 85.68\% & -1.13\% & 86.43\% & -0.26\% & 85.49\% & -1.36\% \\ \hline
        Synonym & 84.55\% & -2.44\% & 85.32\% & -1.55\% & 83.68\% & -3.44\% \\ \hline
        Lemma & 86.50\% & -0.18\% & 86.67\% & 0.01\% & 85.75\% & -1.05\% \\ \hline
        Stem & 86.18\% & -0.56\% & 86.56\% & -0.11\% & 85.57\% & -1.25\% \\ \hline
        RCS & 83.53\% & -3.61\% & 85.91\% & -0.87\% & 82.82\% & -4.44\% \\ \hline
        KCS & 83.80\% & -3.30\% & 85.99\% & -0.77\% & 83.09\% & -4.12\% \\ \hline
        CD & 83.84\% & -3.25\% & 85.82\% & -0.97\% & 84.88\% & -2.05\% \\ \hline
        RW & 86.08\% & -0.67\% & 86.69\% & 0.03\% & 85.48\% & -1.37\% \\ \hline
        BT & 82.64\% & -4.63\% & 82.37\% & -4.95\% & 81.15\% & -6.36\% \\ \hline
        Paraphrase & 84.36\% & -2.65\% & 84.66\% & -2.30\% & 83.21\% & -3.99\% \\ \hline
        Average & 84.72\% & -2.24\% & 85.64\% & -1.17\% & 84.11\% & -2.94\% \\ \hline
        Typos & 83.72\% & -3.39\% & 85.91\% & -0.87\% & 83.60\% & -3.53\% \\ \hline
    \end{tabular}
    \caption{Retrieval accuracy and relative loss across types of noise for unaltered (Regular),  Post Training Contrastive Alignment (CAPOT), and Post Training Contrastive Alignment (CAPOT) using Noisy-ORCAS on TriviaQA dataset with the recall set the size of 200}
    \label{tab:capot-orcasvsnot-200}
\end{table*}

%% file: appendix-capot.tex
\begin{table*}[!ht]
    \centering
    \begin{tabular}{|l|l|l|l|l|l|l|l|l|}
    \hline
        Noise & Regular & Loss & PT & Loss & DA & Loss & CAPOT & Loss \\ \hline
        None & 79.73\% & -0.04\% & 75.04\% & -5.88\% & 79.61\% & -0.15\% & 77.84\% & -2.37\% \\ \hline
        Determiner & 74.59\% & -6.49\% & 72.33\% & -9.29\% & 77.67\% & -2.58\% & 76.23\% & -4.39\% \\ \hline
        Synonym & 68.47\% & -14.16\% & 67.12\% & -15.81\% & 73.07\% & -8.35\% & 71.66\% & -10.12\% \\ \hline
        Lemma & 74.33\% & -6.82\% & 74.79\% & -6.19\% & 77.95\% & -2.23\% & 77.70\% & -2.55\% \\ \hline
        Stem & 74.24\% & -6.92\% & 71.36\% & -10.50\% & 77.95\% & -2.23\% & 76.84\% & -3.62\% \\ \hline
        RCS & 67.12\% & -15.84\% & 66.81\% & -16.20\% & 75.24\% & -5.64\% & 75.43\% & -5.39\% \\ \hline
        KCS & 67.01\% & -15.98\% & 67.26\% & -15.64\% & 75.82\% & -4.91\% & 75.60\% & -5.19\% \\ \hline
        CD & 69.35\% & -13.06\% & 67.48\% & -15.37\% & 75.76\% & -4.98\% & 75.54\% & -5.26\% \\ \hline
        RW & 76.46\% & -4.14\% & 73.68\% & -7.58\% & 78.39\% & -1.68\% & 77.98\% & -2.20\% \\ \hline
        BT & 72.03\% & -9.70\% & 67.23\% & -15.68\% & 72.91\% & -8.56\% & 71.27\% & -10.61\% \\ \hline
        Paraphrase & 72.00\% & -9.73\% & 66.45\% & -16.65\% & 73.05\% & -8.38\% & 71.63\% & -10.15\% \\ \hline
        Average & 71.56\% & -10.28\% & 69.45\% & -12.89\% & 75.78\% & -4.95\% & 74.99\% & -5.95\% \\ \hline
        Typos & 67.83\% & -14.96\% & 67.18\% & -15.74\% & 75.60\% & -5.17\% & 75.52\% & -5.28\% \\ \hline
    \end{tabular}
    \caption{Retrieval accuracy and relative loss across types of noise for unaltered (Regular), PreTrained Alignment (PT),  Data Augmentation (DA), and Post Training Contrastive Alignment (CAPOT) on NQ dataset with the recall set the size of 20}
    \label{tab:capot-nq-20}
\end{table*}
\begin{table*}[!ht]
    \centering
    \begin{tabular}{|l|l|l|l|l|l|l|l|l|}
    \hline
        Noise & Regular & Loss & PT & Loss & DA & Loss & CAPOT & Loss \\ \hline
        None & 85.82\% & -0.19\% & 84.60\% & -1.61\% & 86.29\% & 0.36\% & 85.04\% & -1.09\% \\ \hline
        Determiner & 83.43\% & -2.96\% & 81.91\% & -4.73\% & 84.79\% & -1.38\% & 83.66\% & -2.70\% \\ \hline
        Synonym & 79.75\% & -7.25\% & 73.12\% & -14.95\% & 81.94\% & -4.70\% & 81.08\% & -5.70\% \\ \hline
        Lemma & 82.96\% & -3.51\% & 84.27\% & -1.99\% & 85.32\% & -0.77\% & 84.96\% & -1.19\% \\ \hline
        Stem & 82.96\% & -3.51\% & 82.08\% & -4.54\% & 85.32\% & -0.77\% & 84.68\% & -1.51\% \\ \hline
        RCS & 78.78\% & -8.37\% & 78.95\% & -8.18\% & 83.93\% & -2.38\% & 83.49\% & -2.90\% \\ \hline
        KCS & 78.03\% & -9.24\% & 79.11\% & -7.99\% & 83.55\% & -2.83\% & 83.32\% & -3.09\% \\ \hline
        CD & 79.64\% & -7.37\% & 79.28\% & -7.79\% & 84.29\% & -1.96\% & 83.74\% & -2.61\% \\ \hline
        RW & 85.04\% & -1.09\% & 83.74\% & -2.61\% & 85.51\% & -0.54\% & 85.07\% & -1.06\% \\ \hline
        BT & 80.11\% & -6.83\% & 77.98\% & -9.31\% & 80.53\% & -6.34\% & 79.31\% & -7.76\% \\ \hline
        Paraphrase & 81.30\% & -5.44\% & 78.78\% & -8.37\% & 81.69\% & -4.99\% & 80.78\% & -6.05\% \\ \hline
        Average & 81.62\% & -5.07\% & 79.92\% & -7.05\% & 83.69\% & -2.67\% & 83.01\% & -3.46\% \\ \hline
        Typos & 78.82\% & -8.33\% & 79.11\% & -7.99\% & 83.92\% & -2.39\% & 83.52\% & -2.86\% \\ \hline
    \end{tabular}
    \caption{Retrieval accuracy and relative loss across types of noise for unaltered (Regular), PreTrained Alignment (PT),  Data Augmentation (DA), and Post Training Contrastive Alignment (CAPOT) on NQ dataset with the recall set the size of 100}
    \label{tab:capot-nq-100}
\end{table*}
\begin{table*}[!ht]
    \centering
    \begin{tabular}{|l|l|l|l|l|l|l|l|l|}
    \hline
        Noise & Regular & Loss & PT & Loss & DA & Loss & CAPOT & Loss \\ \hline
        None & 88.25\% & 0.00\% & 87.01\% & -1.41\% & 86.29\% & -2.22\% & 87.23\% & -1.16\% \\ \hline
        Determiner & 85.99\% & -2.56\% & 84.88\% & -3.82\% & 86.73\% & -1.72\% & 86.43\% & -2.07\% \\ \hline
        Synonym & 82.83\% & -6.15\% & 82.11\% & -6.96\% & 84.65\% & -4.08\% & 84.07\% & -4.73\% \\ \hline
        Lemma & 85.13\% & -3.54\% & 86.48\% & -2.00\% & 87.37\% & -1.00\% & 87.23\% & -1.16\% \\ \hline
        Stem & 85.72\% & -2.87\% & 84.88\% & -3.82\% & 87.37\% & -1.00\% & 87.04\% & -1.38\% \\ \hline
        RCS & 81.96\% & -7.13\% & 82.44\% & -6.59\% & 85.96\% & -2.60\% & 85.71\% & -2.88\% \\ \hline
        KCS & 81.52\% & -7.63\% & 82.52\% & -6.49\% & 85.82\% & -2.76\% & 86.12\% & -2.41\% \\ \hline
        CD & 82.04\% & -7.04\% & 82.41\% & -6.62\% & 86.23\% & -2.29\% & 86.65\% & -1.82\% \\ \hline
        RW & 87.34\% & -1.03\% & 86.37\% & -2.13\% & 87.40\% & -0.97\% & 87.51\% & -0.84\% \\ \hline
        BT & 82.54\% & -6.47\% & 80.64\% & -8.63\% & 82.71\% & -6.27\% & 82.05\% & -7.03\% \\ \hline
        Paraphrase & 84.09\% & -4.72\% & 82.19\% & -6.87\% & 83.91\% & -4.92\% & 83.77\% & -5.08\% \\ \hline
        Average & 83.91\% & -4.91\% & 83.49\% & -5.39\% & 85.81\% & -2.76\% & 85.66\% & -2.94\% \\ \hline
        Typos & 81.84\% & -7.27\% & 82.46\% & -6.57\% & 86.00\% & -2.55\% & 86.16\% & -2.37\% \\ \hline
    \end{tabular}
    \caption{Retrieval accuracy and relative loss across types of noise for unaltered (Regular), PreTrained Alignment (PT),  Data Augmentation (DA), and Post Training Contrastive Alignment (CAPOT) on NQ dataset with the recall set the size of 200}
    \label{tab:capot-nq-200}
\end{table*}
\begin{table*}[!ht]
    \centering
    \begin{tabular}{|l|l|l|l|l|l|l|l|l|}
    \hline
        Noise & Regular & Loss & PT & Loss & DA & Loss & CAPOT & Loss \\ \hline
        None & 79.40\% & 0.00\% & 73.52\% & -7.41\% & 76.16\% & -4.08\% & 78.53\% & -1.10\% \\ \hline
        Determiner & 77.36\% & -2.57\% & 71.54\% & -9.90\% & 74.37\% & -6.34\% & 77.76\% & -2.07\% \\ \hline
        Synonym & 75.11\% & -5.41\% & 69.17\% & -12.89\% & 73.22\% & -7.79\% & 75.97\% & -4.33\% \\ \hline
        Lemma & 79.06\% & -0.43\% & 73.41\% & -7.54\% & 76.14\% & -4.10\% & 78.69\% & -0.90\% \\ \hline
        Stem & 78.21\% & -1.50\% & 72.57\% & -8.60\% & 75.74\% & -4.60\% & 78.24\% & -1.46\% \\ \hline
        RCS & 72.64\% & -8.51\% & 67.60\% & -14.86\% & 73.12\% & -7.91\% & 76.74\% & -3.35\% \\ \hline
        KCS & 72.34\% & -8.89\% & 67.87\% & -14.52\% & 73.40\% & -7.55\% & 76.90\% & -3.15\% \\ \hline
        CD & 73.14\% & -7.88\% & 67.86\% & -14.53\% & 73.31\% & -7.68\% & 76.47\% & -3.69\% \\ \hline
        RW & 78.05\% & -1.70\% & 72.80\% & -8.31\% & 75.23\% & -5.25\% & 78.74\% & -0.83\% \\ \hline
        BT & 74.05\% & -6.74\% & 68.13\% & -14.19\% & 70.65\% & -11.02\% & 73.56\% & -7.35\% \\ \hline
        Paraphrase & 75.10\% & -5.41\% & 68.66\% & -13.52\% & 71.65\% & -9.76\% & 74.19\% & -6.56\% \\ \hline
        Average & 75.51\% & -4.90\% & 69.96\% & -11.89\% & 73.68\% & -7.20\% & 76.73\% & -3.37\% \\ \hline
        Typos & 72.71\% & -8.43\% & 67.78\% & -14.64\% & 73.28\% & -7.71\% & 76.71\% & -3.39\% \\ \hline
    \end{tabular}
    \caption{Retrieval accuracy and relative loss across types of noise for unaltered (Regular), PreTrained Alignment (PT),  Data Augmentation (DA), and Post Training Contrastive Alignment (CAPOT) on TriviaQA dataset with the recall set the size of 20}
    \label{tab:capot-trivia-20}
\end{table*}
\begin{table*}[!ht]
    \centering
    
    \begin{tabular}{|l|l|l|l|l|l|l|l|l|}
    \hline
        Noise & Regular & Loss & PT & Loss & DA & Loss & CAPOT & Loss \\ \hline
        None & 84.88\% & -0.15\% & 81.70\% & -3.89\% & 82.93\% & -2.45\% & 84.85\% & -0.19\% \\ \hline
        Determiner & 83.70\% & -1.54\% & 80.49\% & -5.32\% & 81.71\% & -3.88\% & 84.25\% & -0.90\% \\ \hline
        Synonym & 82.27\% & -3.23\% & 78.51\% & -7.64\% & 80.78\% & -4.97\% & 83.12\% & -2.23\% \\ \hline
        Lemma & 84.68\% & -0.39\% & 81.67\% & -3.93\% & 82.85\% & -2.54\% & 84.93\% & -0.10\% \\ \hline
        Stem & 84.20\% & -0.95\% & 81.09\% & -4.61\% & 82.67\% & -2.75\% & 84.82\% & -0.22\% \\ \hline
        RCS & 80.92\% & -4.81\% & 77.83\% & -8.45\% & 81.23\% & -4.44\% & 83.72\% & -1.52\% \\ \hline
        KCS & 81.19\% & -4.49\% & 77.88\% & -8.39\% & 81.29\% & -4.38\% & 83.95\% & -1.25\% \\ \hline
        CD & 81.30\% & -4.37\% & 78.22\% & -7.99\% & 81.19\% & -4.49\% & 83.75\% & -1.48\% \\ \hline
        RW & 84.20\% & -0.95\% & 81.17\% & -4.51\% & 82.56\% & -2.88\% & 84.86\% & -0.18\% \\ \hline
        BT & 80.12\% & -5.75\% & 76.72\% & -9.76\% & 78.24\% & -7.97\% & 80.16\% & -5.70\% \\ \hline
        Paraphrase & 82.21\% & -3.30\% & 78.43\% & -7.74\% & 79.80\% & -6.13\% & 82.29\% & -3.20\% \\ \hline
        Average & 82.48\% & -2.98\% & 79.20\% & -6.83\% & 81.23\% & -4.44\% & 83.58\% & -1.68\% \\ \hline
        Typos & 81.14\% & -4.56\% & 77.98\% & -8.28\% & 81.24\% & -4.44\% & 83.81\% & -1.42\% \\ \hline
    \end{tabular}
    \caption{Retrieval accuracy and relative loss across types of noise for unaltered (Regular), PreTrained Alignment (PT),  Data Augmentation (DA), and Post Training Contrastive Alignment (CAPOT) on TriviaQA dataset with the recall set the size of 100}
    \label{tab:capot-trivia-100}
\end{table*}
\begin{table*}[!ht]
    \centering
    \begin{tabular}{|l|l|l|l|l|l|l|l|l|}
    \hline
        Noise & Regular & Loss & PT & Loss & DA & Loss & CAPOT & Loss \\ \hline
        None & 86.66\% & 0.00\% & 84.03\% & -3.04\% & 84.96\% & -1.97\% & 86.60\% & -0.07\% \\ \hline
        Determiner & 85.68\% & -1.13\% & 83.14\% & -4.06\% & 84.13\% & -2.92\% & 86.43\% & -0.26\% \\ \hline
        Synonym & 84.55\% & -2.44\% & 81.43\% & -6.04\% & 83.18\% & -4.02\% & 85.32\% & -1.55\% \\ \hline
        Lemma & 86.50\% & -0.18\% & 83.97\% & -3.10\% & 84.89\% & -2.04\% & 86.67\% & 0.01\% \\ \hline
        Stem & 86.18\% & -0.56\% & 83.59\% & -3.55\% & 84.77\% & -2.18\% & 86.56\% & -0.11\% \\ \hline
        RCS & 83.53\% & -3.61\% & 81.13\% & -6.38\% & 83.52\% & -3.62\% & 85.91\% & -0.87\% \\ \hline
        KCS & 83.80\% & -3.30\% & 81.18\% & -6.32\% & 83.70\% & -3.42\% & 85.99\% & -0.77\% \\ \hline
        CD & 83.84\% & -3.25\% & 81.30\% & -6.19\% & 83.60\% & -3.53\% & 85.82\% & -0.97\% \\ \hline
        RW & 86.08\% & -0.67\% & 83.74\% & -3.37\% & 84.57\% & -2.42\% & 86.69\% & 0.03\% \\ \hline
        BT & 82.64\% & -4.63\% & 79.41\% & -8.36\% & 80.60\% & -7.00\% & 82.37\% & -4.95\% \\ \hline
        Paraphrase & 84.36\% & -2.65\% & 81.45\% & -6.01\% & 82.66\% & -4.62\% & 84.66\% & -2.30\% \\ \hline
        Average & 84.72\% & -2.24\% & 82.03\% & -5.34\% & 83.56\% & -3.57\% & 85.64\% & -1.17\% \\ \hline
        Typos & 83.72\% & -3.39\% & 81.20\% & -5.47\% & 83.61\% & -3.64\% & 85.91\% & -0.87\% \\ \hline
    \end{tabular}
    \caption{Retrieval accuracy and relative loss across types of noise for unaltered (Regular), PreTrained Alignment (PT),  Data Augmentation (DA), and Post Training Contrastive Alignment (CAPOT) on TriviaQA dataset with the recall set the size of 200}
    \label{tab:capot-trivia-200}
\end{table*}
\begin{table*}[!ht]
    \centering
   
    \begin{tabular}{|l|l|l|l|l|l|l|l|l|}
    \hline
        Noise & Regular & Loss & PT & Loss & DA & Loss & CAPOT & Loss \\ \hline
        None & 32.39\% & 0.00\% & 19.05\% & -41.18\% & 20.70\% & -35.89\% & 25.38\% & -21.39\% \\ \hline
        Determiner & 25.85\% & -20.20\% & 15.50\% & -52.15\% & 16.42\% & -49.15\% & 25.25\% & -21.80\% \\ \hline
        Synonym & 20.63\% & -36.30\% & 11.56\% & -64.30\% & 13.56\% & -58.01\% & 17.92\% & -44.49\% \\ \hline
        Lemma & 31.50\% & -2.74\% & 18.76\% & -42.07\% & 20.66\% & -36.03\% & 25.90\% & -19.80\% \\ \hline
        Stem & 26.87\% & -17.04\% & 16.72\% & -48.39\% & 18.68\% & -42.14\% & 25.13\% & -22.16\% \\ \hline
        RCS & 16.85\% & -47.96\% & 12.09\% & -62.66\% & 15.13\% & -53.16\% & 22.91\% & -29.05\% \\ \hline
        KCS & 16.00\% & -50.62\% & 12.16\% & -62.46\% & 14.86\% & -53.97\% & 22.63\% & -29.91\% \\ \hline
        CD & 18.00\% & -44.41\% & 12.52\% & -61.36\% & 15.02\% & -53.49\% & 22.46\% & -30.45\% \\ \hline
        RW & 31.19\% & -3.70\% & 18.41\% & -43.17\% & 19.91\% & -38.35\% & 28.18\% & -12.72\% \\ \hline
        BT & 26.76\% & -17.40\% & 15.52\% & -52.08\% & 16.91\% & -47.63\% & 20.80\% & -35.58\% \\ \hline
        Paraphrase & 26.56\% & -17.99\% & 15.51\% & -52.13\% & 16.82\% & -47.90\% & 21.05\% & -34.79\% \\ \hline
        Average & 24.02\% & -25.84\% & 14.87\% & -54.08\% & 16.80\% & -47.98\% & 23.22\% & -28.08\% \\ \hline
        Typos & 16.95\% & -47.66\% & 12.26\% & -62.16\% & 15.00\% & -53.54\% & 22.67\% & -29.80\% \\ \hline
    \end{tabular}
    \caption{MRR@10 and relative loss across types of noise for unaltered (Regular), PreTrained Alignment (PT),  Data Augmentation (DA), and Post Training Contrastive Alignment (CAPOT) on MSMARCO dataset.}
    \label{tab:capot-msmarco-mrr10}
\end{table*}
\begin{table*}[!ht]
    \centering
    
    \begin{tabular}{|l|l|l|l|l|l|l|l|l|}
    \hline
        Noise & Regular & Loss & PT & Loss & DA & Loss & CAPOT & Loss \\ \hline
        None & 32.39\% & 0.00\% & 19.05\% & -41.18\% & 20.70\% & -35.89\% & 25.38\% & -21.39\% \\ \hline
        Determiner & 25.85\% & -20.20\% & 15.50\% & -52.15\% & 16.42\% & -49.15\% & 25.25\% & -21.80\% \\ \hline
        Synonym & 20.63\% & -36.30\% & 11.56\% & -64.30\% & 13.56\% & -58.01\% & 17.92\% & -44.49\% \\ \hline
        Lemma & 31.50\% & -2.74\% & 18.76\% & -42.07\% & 20.66\% & -36.03\% & 25.90\% & -19.80\% \\ \hline
        Stem & 26.87\% & -17.04\% & 16.72\% & -48.39\% & 18.68\% & -42.14\% & 25.13\% & -22.16\% \\ \hline
        RCS & 16.85\% & -47.96\% & 12.09\% & -62.66\% & 15.13\% & -53.16\% & 22.91\% & -29.05\% \\ \hline
        KCS & 16.00\% & -50.62\% & 12.16\% & -62.46\% & 14.86\% & -53.97\% & 22.63\% & -29.91\% \\ \hline
        CD & 18.00\% & -44.41\% & 12.52\% & -61.36\% & 15.02\% & -53.49\% & 22.46\% & -30.45\% \\ \hline
        RW & 31.19\% & -3.70\% & 18.41\% & -43.17\% & 19.91\% & -38.35\% & 28.18\% & -12.72\% \\ \hline
        BT & 26.76\% & -17.40\% & 15.52\% & -52.08\% & 16.91\% & -47.63\% & 20.80\% & -35.58\% \\ \hline
        Paraphrase & 26.56\% & -17.99\% & 15.51\% & -52.13\% & 16.82\% & -47.90\% & 21.05\% & -34.79\% \\ \hline
        Average & 24.02\% & -25.84\% & 14.87\% & -54.08\% & 16.80\% & -47.98\% & 23.22\% & -28.08\% \\ \hline
        Typos & 16.95\% & -47.66\% & 12.26\% & -62.16\% & 15.00\% & -53.54\% & 22.67\% & -29.80\% \\ \hline
    \end{tabular}
    \caption{Retrieval accuracy and relative loss across types of noise for unaltered (Regular), PreTrained Alignment (PT),  Data Augmentation (DA), and Post Training Contrastive Alignment (CAPOT) on MSMARCO dataset with the recall set the size of 20}
    \label{tab:capot-msmarco-20}
\end{table*}
\begin{table*}[!ht]
    \centering
    
    \begin{tabular}{|l|l|l|l|l|l|l|l|l|}
    \hline
        Noise & Regular & Loss & PT & Loss & DA & Loss & CAPOT & Loss \\ \hline
        None & 88.79\% & 0.00\% & 67.87\% & -23.57\% & 71.83\% & -19.10\% & 79.70\% & -10.24\% \\ \hline
        Determiner & 77.66\% & -12.54\% & 58.57\% & -34.04\% & 63.02\% & -29.02\% & 76.99\% & -13.29\% \\ \hline
        Synonym & 68.09\% & -23.31\% & 47.12\% & -46.93\% & 55.01\% & -38.04\% & 61.96\% & -30.21\% \\ \hline
        Lemma & 87.54\% & -1.41\% & 67.45\% & -24.03\% & 71.73\% & -19.21\% & 80.87\% & -8.92\% \\ \hline
        Stem & 79.30\% & -10.69\% & 62.36\% & -29.76\% & 68.05\% & -23.36\% & 79.23\% & -10.77\% \\ \hline
        RCS & 58.00\% & -34.68\% & 47.56\% & -46.43\% & 58.81\% & -33.76\% & 72.36\% & -18.50\% \\ \hline
        KCS & 56.09\% & -36.83\% & 47.69\% & -46.29\% & 59.03\% & -33.52\% & 72.99\% & -17.79\% \\ \hline
        CD & 61.97\% & -30.21\% & 49.99\% & -43.70\% & 58.77\% & -33.81\% & 71.99\% & -18.92\% \\ \hline
        RW & 87.57\% & -1.37\% & 66.89\% & -24.66\% & 70.74\% & -20.32\% & 83.48\% & -5.98\% \\ \hline
        BT & 78.60\% & -11.47\% & 57.97\% & -34.72\% & 62.49\% & -29.62\% & 68.91\% & -22.39\% \\ \hline
        Paraphrase & 79.67\% & -10.27\% & 59.04\% & -33.51\% & 63.74\% & -28.21\% & 70.57\% & -20.52\% \\ \hline
        Average & 74.84\% & -15.71\% & 56.46\% & -36.41\% & 63.14\% & -28.89\% & 73.94\% & -16.73\% \\ \hline
        Typos & 58.68\% & -33.91\% & 48.41\% & -45.47\% & 58.87\% & -33.70\% & 72.45\% & -18.40\% \\ \hline
    \end{tabular}
    \caption{Retrieval accuracy and relative loss across types of noise for unaltered (Regular), PreTrained Alignment (PT),  Data Augmentation (DA), and Post Training Contrastive Alignment (CAPOT) on MSMARCO dataset with the recall set the size of 100}
    \label{tab:capot-msmarco-100}
\end{table*}
\begin{table*}[!ht]
    \centering
    
    \begin{tabular}{|l|l|l|l|l|l|l|l|l|}
    \hline
        Noise & Regular & Loss & PT & Loss & DA & Loss & CAPOT & Loss \\ \hline
        None & 93.78\% & 0.00\% & 74.99\% & -20.04\% & 79.57\% & -15.15\% & 85.69\% & -8.63\% \\ \hline
        Determiner & 83.56\% & -10.89\% & 65.60\% & -30.05\% & 70.56\% & -24.76\% & 82.68\% & -11.84\% \\ \hline
        Synonym & 74.54\% & -20.52\% & 54.54\% & -41.84\% & 62.58\% & -33.27\% & 68.72\% & -26.72\% \\ \hline
        Lemma & 92.57\% & -1.29\% & 74.61\% & -20.44\% & 79.31\% & -15.43\% & 86.68\% & -7.57\% \\ \hline
        Stem & 84.70\% & -9.69\% & 69.81\% & -25.56\% & 75.62\% & -19.37\% & 85.24\% & -9.10\% \\ \hline
        RCS & 64.42\% & -31.31\% & 54.66\% & -41.72\% & 66.40\% & -29.19\% & 78.87\% & -15.90\% \\ \hline
        KCS & 62.87\% & -32.96\% & 54.67\% & -41.70\% & 67.32\% & -28.21\% & 79.48\% & -15.24\% \\ \hline
        CD & 68.36\% & -27.11\% & 56.81\% & -39.43\% & 66.91\% & -28.66\% & 78.11\% & -16.71\% \\ \hline
        RW & 92.63\% & -1.23\% & 74.27\% & -20.80\% & 78.04\% & -16.79\% & 89.43\% & -4.64\% \\ \hline
        BT & 84.08\% & -10.34\% & 65.56\% & -30.09\% & 70.44\% & -24.88\% & 75.27\% & -19.74\% \\ \hline
        Paraphrase & 85.82\% & -8.49\% & 67.15\% & -28.40\% & 71.35\% & -23.92\% & 77.48\% & -17.38\% \\ \hline
        Average & 80.67\% & -13.98\% & 63.77\% & -32.00\% & 70.85\% & -24.45\% & 80.20\% & -14.48\% \\ \hline
        Typos & 65.22\% & -30.46\% & 55.38\% & -40.95\% & 66.88\% & -28.69\% & 78.82\% & -15.95\% \\ \hline
    \end{tabular}
    \caption{Retrieval accuracy and relative loss across types of noise for unaltered (Regular), PreTrained Alignment (PT),  Data Augmentation (DA), and Post Training Contrastive Alignment (CAPOT) on MSMARCO dataset with the recall set the size of 200}
    \label{tab:capot-msmarco-200}
\end{table*}